\documentclass[prl,twocolumn,superscriptaddress,nobibnotes,letterpaper]{revtex4}

\usepackage[pdftex]{graphicx}
\usepackage[pdftex]{epsfig}
\usepackage{amsmath}
\usepackage{amssymb}
\usepackage{amsfonts}
\usepackage[usenames, dvipsnames]{color}
\usepackage{wrapfig}
\usepackage{eucal}
\usepackage{hhline}
\usepackage{threeparttable}
\usepackage{supertabular}
\usepackage{multirow}
\usepackage{tabularx}

\def\be{\begin{equation}}
\def\ee{\end{equation}}
\def\bea{\begin{eqnarray}}
\def\eea{\end{eqnarray}}

\newcolumntype{Y}{>{\centering\arraybackslash}X}

\begin{document}

\pagenumbering{arabic}

\title{\fontsize{11.4pt}{10pt}\selectfont Mechanical Resonators for Quantum Optomechanics Experiments at Room Temperature}\thanks{This work was published in Phys.\ Rev.\ Lett.\ \textbf{116}, 147202 (2016).}

\normalfont

\author{R.\ A.\ Norte}
\affiliation{Kavli Institute of Nanoscience, Delft University of Technology, Lorentzweg 1, 2628CJ Delft, The Netherlands}
\author{J.\ P.\ Moura}
\affiliation{Kavli Institute of Nanoscience, Delft University of Technology, Lorentzweg 1, 2628CJ Delft, The Netherlands}
\author{S.\ Gr\"oblacher}
\email{s.groeblacher@tudelft.nl}
\affiliation{Kavli Institute of Nanoscience, Delft University of Technology, Lorentzweg 1, 2628CJ Delft, The Netherlands}


\begin{abstract}
All quantum optomechanics experiments to date operate at cryogenic temperatures, imposing severe technical challenges and fundamental constraints. Here we present a novel design of on-chip mechanical resonators which exhibit fundamental modes with frequencies $f$ and mechanical quality factors $Q_\mathrm{m}$ sufficient to enter the optomechanical quantum regime at room temperature. We overcome previous limitations by designing ultrathin, high-stress silicon nitride (Si$_3$N$_4$) membranes, with tensile stress in the resonators' clamps close to the ultimate yield strength of the material. By patterning a photonic crystal on the SiN membranes, we observe reflectivities greater than 99\%. These on-chip resonators have remarkably low mechanical dissipation, with $Q_\mathrm{m}$$\sim$$10^8$, while at the same time exhibiting large reflectivities. This makes them a unique platform for experiments towards the observation of massive quantum behavior at room temperature.
\end{abstract}

\maketitle

Optomechanical systems, where light is coupled to mechanical motion via the radiation pressure force, have generated enormous interest over the past years. They are promising candidates for testing macroscopic quantum physics, have great potential as quantum transducers between distinct quantum systems and have unique capabilities for sensing applications~\cite{Aspelmeyer2014}. State-of-the-art systems have recently demonstrated ground state cooling~\cite{Chan2011,Teufel2011b}, mechanical quantum state preparation~\cite{OConnell2010,Riedinger2016}, entanglement~\cite{Palomaki2013} and squeezing of both the optical~\cite{Safavi-Naeini2013b,Purdy2013b} as well as the mechanical mode~\cite{wollman2015quantum,Pirkkalainen2015,Lecocq2015}. Experiments involving such optomechanical systems in the quantum regime are technically very challenging and so far have exclusively operated at cryogenic temperatures. This poses serious restrictions on the type of experiments that are feasible. Without the need for cryogenic precooling, one could envision their use as hybrid quantum systems with, for example, atomic gases~\cite{Treutlein2007} or single atoms~\cite{Hammerer2009b}. It would also open up practical avenues for real world applications of such quantum optomechanical systems. 

One of the most successful implementation of mechanical oscillators for such (quantum) optomechanics experiments are devices made of high-stress silicon nitride (Si$_3$N$_4$), which have been utilized in quantum-limited accelerometers~\cite{krause2012high}, coupling of their motion to ultracold atoms~\cite{jockel2014sympathetic,Stamper-Kurn2014}, optomechanics in 3D microwave cavities~\cite{yuan2015large}, microwave-to-optical wavelength conversion~\cite{andrews2014bidirectional}, and quadratic coupling in cavity optomechanics~\cite{thompson2008strong}.

However, even these low mechanical dissipation oscillators have to date not operated in a regime where realistic quantum experiments at room temperature are feasible. The benchmark for this elusive regime is the $f \times Q_\mathrm{m}$ product of the resonator which requires the mechanical quality factor $Q_\mathrm{m}$ to be larger than the number of thermal phonons at room temperature ($f \times Q_\mathrm{m} >k_\mathrm{B} T_{\mathrm{room}}/h$), with $h$ being the Planck and $k_\mathrm{B}$ the Boltzmann constant~\cite{Marquardt2007}. This regime will allow for ground state cooling using the radiation pressure force and hence for experiments operating in the quantum domain. In general, the quality factor-frequency product also determines the number of coherent oscillations the resonator can undergo before one phonon enters the system; i.e., $N_{\mathrm{osc}} = Q_\mathrm{m} f\cdot h/2\pi k_\mathrm{B} T$. Experimental realizations of on-chip mechanical resonators that exceed this requirement have all been demonstrated in higher-order mechanical modes~\cite{Wilson2009,Chakram2014}. Such mechanical systems are however in practice not useful for cooling experiments as higher order modes are enveloped by numerous neighboring modes, which increase the displacement background noise as one cools the mode of interest. To avoid this limit, it is important to couple to the fundamental mode of the resonator. An additional challenge is to operate at mechanical frequencies beyond $10^5$~Hz, where commercial lasers exhibit a minimal amount of classical noise and can relatively easily be quantum limited to shot noise in order to avoid heating or decoherence of the mechanics through noise~\cite{Safavi-Naeini2013a}.
Another difficulty for realistic quantum optomechanics experiments at room temperature is that often good mechanical quality is mutually exclusive with good optical reflectivity~\cite{Harry2002,Groeblacher2009a,kleckner2011optomechanical}. This limits the achievable coupling rates and increases the necessary optical power to a level where absorption potentially becomes a practical limitation for cooling and quantum experiments.

\begin{figure}[t]
\begin{center}
\includegraphics[width=.9\columnwidth]{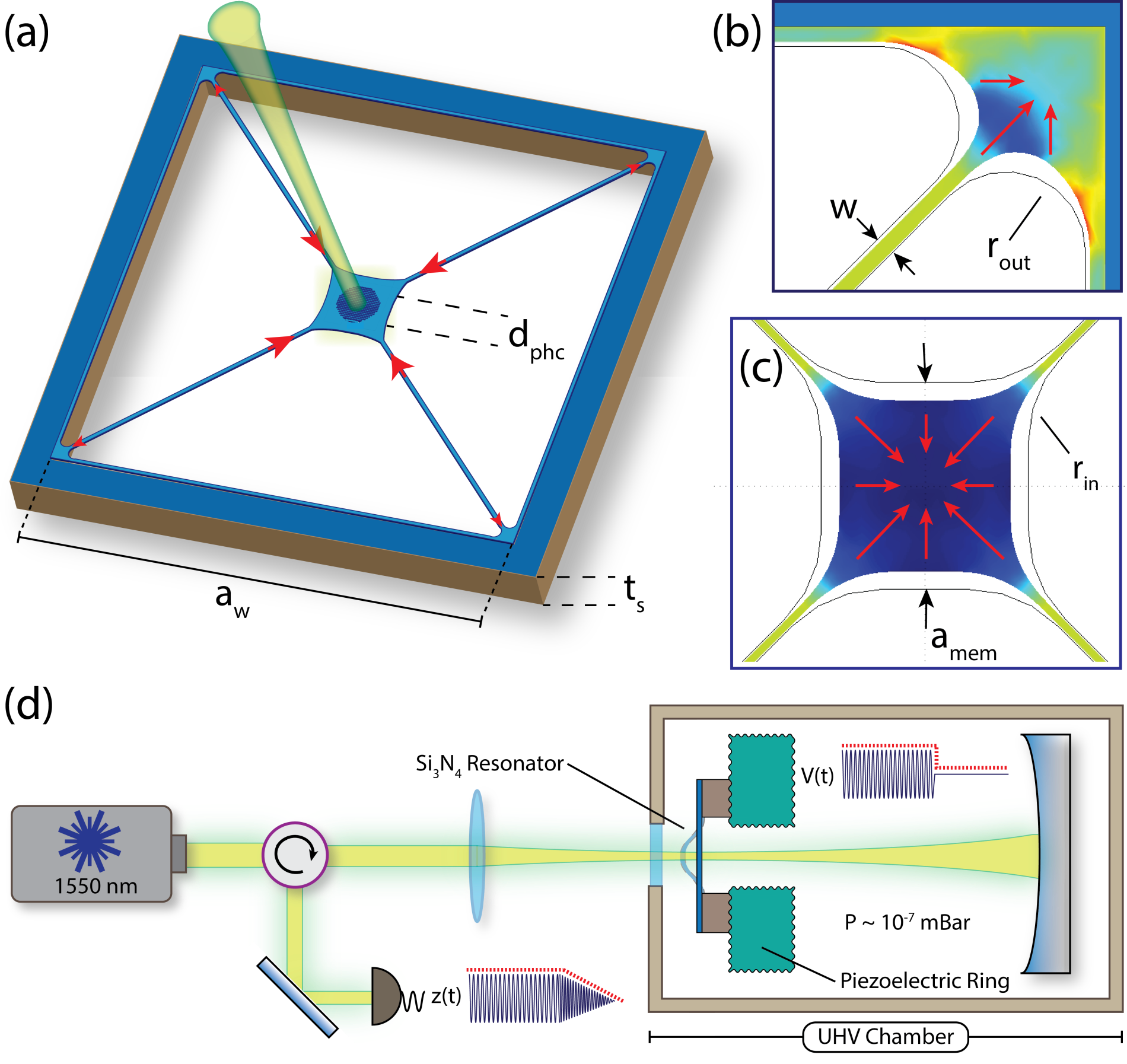}
\caption{(a) Schematic of our tethered membrane which consists of a central membrane connected to the silicon substrate through a series of thin tethers. A central photonic crystal reflector can be used for increased reflectivities. Shown are our design parameters which are individually swept keeping all others constant to observe their effect on the $f \times Q_\mathrm{m}$ product. Finite element simulations map out the strain throughout the resonator and calculate deformation due to the relaxation of the material at the (b) tether clamps and (c) central membrane, which together significantly enhance the tensile stress in the tethers. (d) Schematic of ringdown measurement setup. The membrane is resonantly driven by a piezoelectric transducer inside an optical interferometer. We determine the mechanical dissipation by observing the ringdown of the mechanical resonance (see text for details).}
\label{fig:1}
\end{center} 
\end{figure}

In this Letter, we demonstrate the first optomechanical platform that overcomes all these limitations, paving the way for room temperature experiments in the quantum regime. We fabricate on-chip optomechanical mirrors which exhibit $f \times Q_\mathrm{m}$ products of their fundamental mechanical modes above the requirement for ground state cooling without cryogenics (Fig.~\ref{fig:1}). With a center-of-mass frequency of $\sim$150~kHz and mechanical quality factor $Q_\mathrm{m}$$\sim$$10^8$, this new generation of Si$_3$N$_4$ tethered membranes are on par with the state of the art in optically levitated nanospheres, known for their extreme mechanical isolation and ultralow dissipation, which are only limited by gas-molecule collisions in high vacuum and photon recoil heating~\cite{Chang2010,gieseler2013thermal}. In order to achieve such remarkably low dissipation rates of $\Gamma_\mathrm{m}/2\pi=f/Q_\mathrm{m}\approx 1.4$~mHz with a tethered system, we design ultrathin high-stress Si$_3$N$_4$ membranes which enhance the intrinsic stress in crucial tether regions -- significantly reducing clamping and bending losses~\cite{Zhang2015}. A key observation is that high-stress membranes have mechanical frequencies which are stress dominated, meaning that one can minimize the thickness of the resonator in order to reduce bending losses without significantly reducing the mechanical mode frequencies. We fabricate tethered membranes with ultralow dissipation rates by engineering up to 6~GPa of stress within films as thin as 15~nm and intrinsic stress of 1.3~GPa. Using finite element simulations to calculate the stress throughout the resonators, we push the tensile stress in the resonator's clamps to values near the ultimate yield strength of low-pressure chemical vapor deposition (LPCVD) Si$_3$N$_4$. In addition, we are able to show that we can increase the intrinsic optical reflectivity $R$ of these membranes to up to $99.3$\% by patterning a two-dimensional photonic crystal onto the structure. These devices enable coupling rates that allow ground state cooling from room temperature with realistic parameters in a membrane-in-the-middle design~\cite{nunnenkamp2010cooling,deng2012performance,usami2012optical,jayich2008dispersive} and can also be used as an end mirror of a Fabry-P\'{e}rot cavity~\cite{Groeblacher2009a,Gigan2006,Arcizet2006b,Schliesser2006}.\\

\begin{figure}[t]
\begin{center}
\includegraphics[width=1\columnwidth]{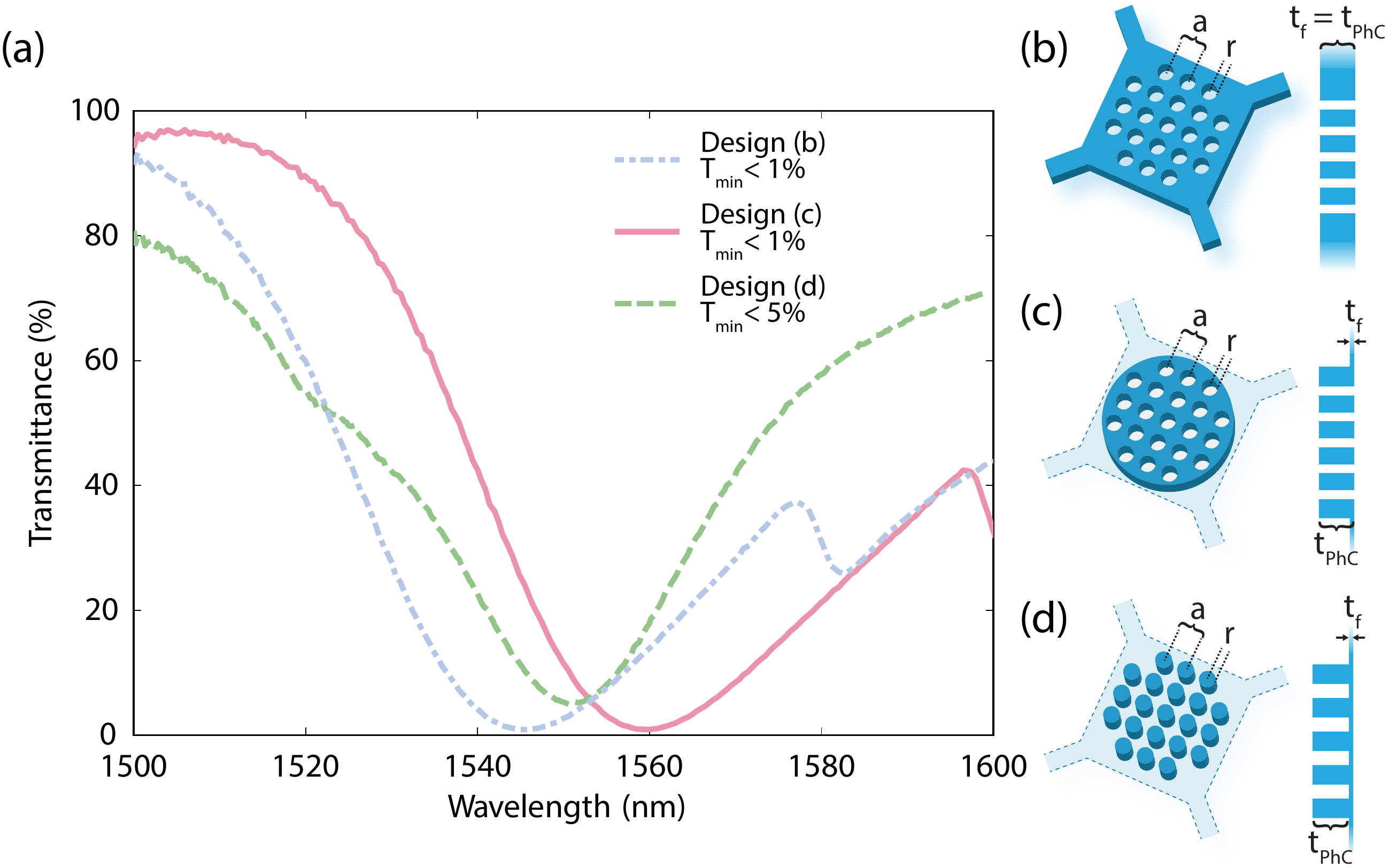}
\caption{(a) The measured transmittance $T$ for three different designs of tethered mechanical membranes (b) -- (d). The thickness of the photonic crystal $t_\mathrm{PhC}$ was chosen to be 200~nm for all designs. The maximum reflectivity ($1-T_\mathrm{min}$) strongly depends on the thickness of the silicon nitride (see SI). For (b), this is in competition with the requirement of thin films (i.e.\ small $t_\mathrm{f}$) for good mechanical quality (cf.\ Fig.~\ref{fig:3}). We overcome this limitation by either leaving a thick cylindrical slab in the center of the membrane (c) or by using an inverse photonic crystal design (d). These designs decouple the mechanical and optical properties of the membrane and show similar optical performances to (b). We use a simple transmittance measurement to infer reflectivity. We experimentally confirm this by using a PhC membrane as one end mirror of a Fabry-P\'erot cavity in combination with a second mirror with known reflectivity and measuring the finesse.}
\label{fig:PhC}
\end{center} 
\end{figure}

The central mirror on the tethered membranes is a 2D photonic crystal device, that is designed using finite difference time domain (FDTD) simulations. They are similar to previous designs of grating reflectors~\cite{Stambaugh2015} and photonic crystals (PhC)~\cite{Bui2012,Makles2015}, which usually consist of an array of either lines or holes etched into the dielectric, respectively. Such a periodic change in the refractive index allows for a band gap to be tailored for a specific wavelength, resulting in (simulated) reflectivities $>$99.9\%. Experiments to date employing such photonic crystal mirrors for optomechanical resonators have however suffered from a trade-off between high reflectivity and high-mechanical quality, due to the requirement of thick dielectrics in order to achieve large $R$~\cite{Makles2015}. We confirm this by measuring PhC mirrors consisting of a square lattice of holes as a function of silicon nitride thickness (see SI). In order to circumvent this design issue we follow two slightly different approaches:\ we either leave a thick cylindrical slab of SiN around the PhC (see Fig.~\ref{fig:PhC}c) or instead of using holes we use pillars, resulting in an inverse photonic crystal (see Fig.~\ref{fig:PhC}d). While the latter design still suffers from fabrication imperfections resulting in reflectivities of $\sim95\%$, the other new design allows us to reach $R>99\%$ (Fig.~\ref{fig:PhC}).
We show that both methods decouple the mechanical from the optical properties and allow for optomechanical devices with simultaneous high $Q_\mathrm{m}$ and high $R$.

We also study the tethered membranes' mechanical parameters, several of which suffer from the conventional trade-off between either good mechanical quality factor or high frequency (see Fig.~\ref{fig:1} and Table~IV, Supplementary Information). For example, extending the length of the tethers by increasing the size of the window $a_\mathrm{w}$ in order to reduce clamping and bending losses not only results in higher mechanical quality factors but also lower frequencies~\cite{verbridge2008size}. Achieving ultrahigh $Q_\mathrm{m}$ with this parameter will result in very low fundamental modes, where low-frequency classical noise in lasers becomes a compelling limitation on radiation-pressure sideband cooling (for more detail and discussion on design parameters see the Supplementary Information).

It is known that for fundamental modes of long, thin nanostrings (width, $w\approx3$--$5~\mathrm{\mu m}$), mechanical dissipation starts to be dominated by bending losses~\cite{schmid2011damping}. By accounting for the dominating effect of flexural bending near the clamps of tensile strings one can express the quality factor for the fundamental mode of doubly clamped strings as

\begin{equation}\label{eq:bendingQ}
Q_{\mathrm{str}} = \left[\frac{\pi^2}{12} \frac{E}{\sigma} \left(\frac{t_\mathrm{f}}{L}\right)^2 + 1.0887 \sqrt{\frac{E}{\sigma}} \left(\frac{t_\mathrm{f}}{L}\right)\right]^{-1} Q_{\mathrm{bending}}
\end{equation}

where $E$ is the Si$_3$N$_4$ Young's modulus, $\sigma$ is the stress in the string, $t_\mathrm{f}$ is the film thickness, $L$ is the length of the nanostring, and $Q_{\mathrm{bending}}$ is the quality factor due to bending losses in a relaxed string, which is mostly dependent on intrinsic material damping. From Eq.~\ref{eq:bendingQ}, one finds that the quality factor of a string can be enhanced by using thinner strings with decreasing $t_\mathrm{f}$. A crucial observation from thin plate theory is that  membranes under large tensile forces have stress-dominated mechanical frequencies (i.e.\ $f_{11} = 1/(\sqrt{2}L)\sqrt{\sigma/\rho}$, where $\rho$ is the material's density) which are independent of membrane thickness. Since our devices' dissipation is dominated by bending losses through the tethers and have mechanical modes which are independent of thickness, we can engineer thinner membranes with increased mechanical quality factors and negligible impact on the frequency. This design scheme overcomes the trade-off between $Q_\mathrm{m}$ and $f$ in order to realize optomechanical resonators with unprecedented enhancement of the fundamental mode $f \times Q_\mathrm{m}$ product.

In addition, substrate thickness plays an important role in anchoring losses for out-of-plane fundamental modes, where larger vibrational displacements of thinner substrates near the clamping points of a fundamental mode significantly increase mechanical dissipation~\cite{darvishian2015effect}. Previous studies found some enhancement in 30--50~nm thick square membranes when moving from 200~$\mu$m substrates to 500~$\mu$m with a negligible effect in thicker Si$_3$N$_4$ films ($t_\mathrm{f}\geq 100$~nm)~\cite{wilson2012cavity}. Silicon substrate thickness and silicon nitride film thickness are the parameters we focus on to achieve significant enhancements in $f \times Q_\mathrm{m}$. We investigate these effects by fabricating our resonators from ultrathin films $\geq$15~nm and on substrates as thick as 900~$\mu$m (for fabrication details see the Supplementary Information).

\begin{figure}[t]
\begin{center}
\includegraphics[width=.9\columnwidth]{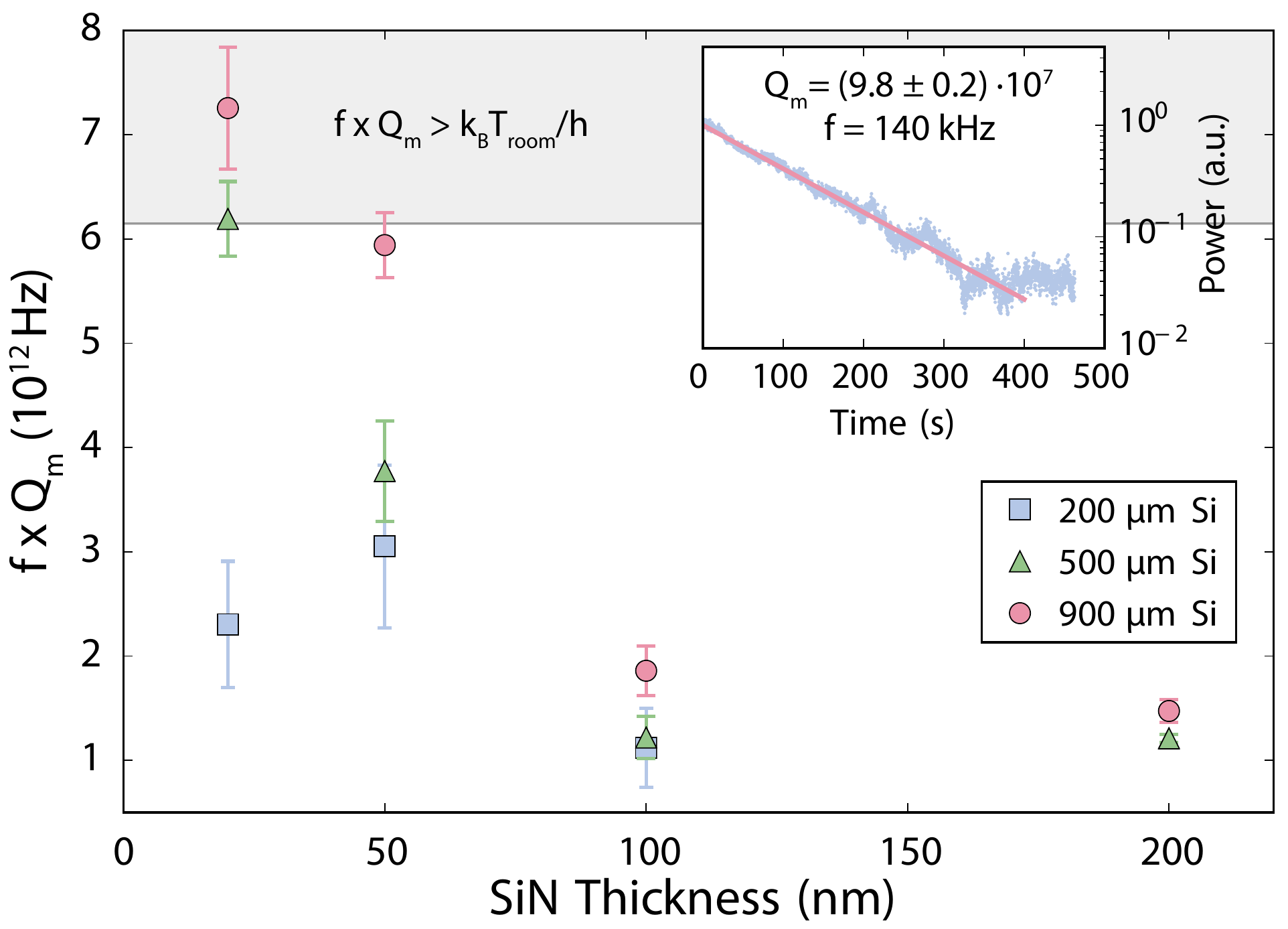}
\caption{Shown is the $f \times Q_\mathrm{m}$ product for the nominal design (see SI) of our tethered membrane for various silicon substrate (\{200, 500, 900\}~$\mu$m) and Si$_3$N$_4$ film (\{20, 50, 100, 200\}~nm) thicknesses. Each point is the average of measurements of resonators with identical geometry. In addition we verified that the results are independent of the particular PhC design. For thin silicon substrates ($t_\mathrm{s} = 200~\mu$m), the anchoring losses completely dominate and result in minimal enhancement of the mechanical quality factor even at $t_\mathrm{f} = 20$~nm. However, these ultrathin films exhibit $f \times Q_\mathrm{m}$ products above the ground state cooling limit (shaded region) when fabricated on thicker substrates, where anchoring losses are less pronounced. The inset shows the ringdown measurement for the best observed value for a single device with $Q_\mathrm{m}=9.8\pm0.2\times10^7$ at $f=140$~kHz for a 20~nm thick film, with otherwise nominal parameters.}
\label{fig:3}
\end{center} 
\end{figure}

We determine the mechanical quality factors of our membranes by performing ringdown measurements using a piezoelectric stack in an optical interferometer (see Figure~\ref{fig:1}d). Due to viscous damping, which becomes increasingly dominant with thinner membranes, our measurements are conducted inside an ultrahigh vacuum chamber at $<$$10^{-7}$~mbar. Positioning stages are used to align the chip with the membranes to a 20~$\mu$m spot of a 1550~nm laser. The chip is placed onto the piezolectric stack under its own weight since any type of clamping or gluing of the chip to the experimental setup can reduce the mechanical quality factor by several orders of magnitude~\cite{wilson2012cavity}. Each of our chips has 9--16 resonators which allow us to collect several data points for each parameter sweep. We start with a nominal resonator design (see Supplementary Information for details) and adjust each parameter keeping all others fixed. In Fig.~\ref{fig:3}, we plot the effects on $f \times Q_\mathrm{m}$ by varying the thickness of the silicon nitride films and silicon substrates for the nominal membrane design. We find that losses in thicker Si$_3$N$_4$ resonators ($t_\mathrm{f} \geq  100$~nm) are dominated by bending  losses. From the data, we also see that low dissipation in ultrathin resonators can be completely dominated by anchoring losses in thin substrates ($\sim$$200~\mu$m). A clear enhancement in the $f \times Q_\mathrm{m}$ product is observed, consistently exceeding the above requirement for ground state cooling at room temperatures as one fabricates thinner tethered membranes on increasingly thicker substrates, where anchoring losses are no longer a limiting factor on $Q_\mathrm{m}$.

\begin{figure}[t]
\begin{center}
\includegraphics[width=.9\columnwidth]{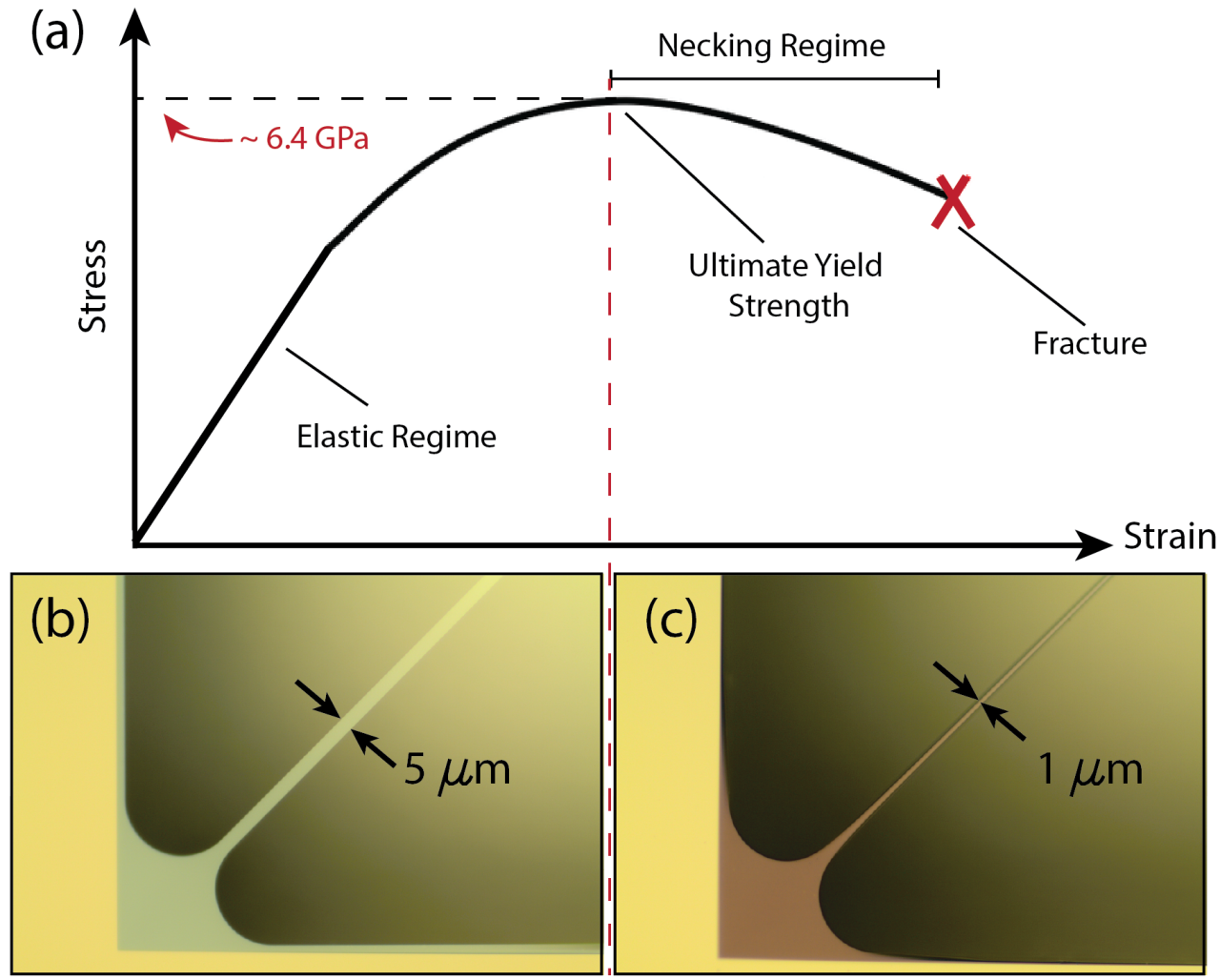}
\caption{(a) Sketch of the stress-strain curve for silicon nitride. We design our structures such that for the thinnest films we operate in the regime close to the ultimate yield strength of the material. (b) Microscope image showing a nominal clamp with outer fillet $r_\mathrm{out} = 20~\mu$m and tether width $w = 5~\mu$m when properly fabricated with turbulence reducing holders. (c) shows typical plastic deformation in very thin films ($\leq$20~nm). This occurs at the tether clamps where the stress is the largest and when the resonators are being exposed to small viscous forces or temperature gradients during fabrication. The tether is deformed to a width of $\sim$1~$\mu$m.} 
\label{fig:4}
\end{center} 
\end{figure}

In order to push the devices to their material limits, we engineer the stress at the clamps (the dominant source of flexural bending loss) to just short of the Si$_3$N$_4$ ultimate yield strength ($\approx 6.4$~GPa~\cite{kaushik2005wafer}). At thicknesses $\leq$20~nm, the silicon nitride membranes become increasingly susceptible to plastic deformation when subjected to small viscous forces due to handling or large temperature variations during wet chemical processing. Fig.~\ref{fig:4}b shows the necking that occurs when the resonator is subjected to these small forces which result in large frequency drops from $\sim$$170$~kHz to $\sim$$60$~kHz and reduction in $Q_\mathrm{m}$ to $\sim$$10^5$. Necking is a form of irreversible plastic deformation that occurs when excessively large strains localize in small cross sections. In order to minimize such effects, we use polytetrafluoroethylene (PTFE) turbulence shielding holders~\cite{norte2014} which gently dilute the resonators from one liquid to the next during fabrication. This significantly reduces the resonators' exposure to viscous forces and surface tension in the wet chemistry processes required to release, clean and rinse the sample thoroughly~\cite{norte2014}. At thicknesses below 20~nm, we observe consistently reduced $Q_\mathrm{m}$$\sim$$10^7$. Ultrathin films below 20~nm produce delicate structures which make it difficult to attribute the lower quality factors to intrinsic limits of the sensitive handling during fabrication or whether other surface-dominant loss mechanisms, such as Akhiezer damping, become more dominant loss channels. The inset in Fig.~\ref{fig:3} shows a ringdown for our best device at 20~nm silicon nitride thickness, 900~$\mu$m Si wafer thickness, and mechanical quality factor $Q_\mathrm{m} = 9.8\pm0.2\times 10^7$ at a fundamental mode frequency of 140~kHz ($f\times Q_\mathrm{m}=1.37\times10^{13}$~Hz).

With the demonstrated combination of large reflectivities and low mechanical dissipation we will be able to access the optomechanical quantum regime from room temperature. A first demonstration of such quantum behavior could be achieved by cooling the mechanical mode into its quantum ground state, which is already realistic with these device parameters (see Supplementary Information for detailed calculations).

To conclude, we studied the dissipation mechanisms in a new regime of ultrathin, highly stressed optomechanical resonators and their effects on mechanical quality factor and fundamental mode frequency. By moving to these thin devices on thick silicon substrates and by engineering the stress in their tethers to near the ultimate yield strength of Si$_3$N$_4$, we are able to overcome a well-known trade-off between frequency and mechanical quality factor to achieve fundamental modes $f \times Q_\mathrm{m}$, which are more than twice the requirement for quantum ground state cooling from room temperature. Our on-chip device performances are on par with the best values measured for optically levitated nanospheres without the need for high power trapping lasers or complex experimental setups. 
We also demonstrate the possibility to combine our resonators with different photonic crystal mirror designs which result in reflectivities between $95\%$ and $99\%$, while simultaneously achieving ultrahigh $Q_\mathrm{m}$. These results allow us to finally realize experiments to laser cool a mechanical oscillator from room temperature to its quantum ground state (see SI).
Such reflective tethered membranes are also ideal for optical trapping configurations that enhance the frequency and the mechanical quality factor even further, while avoiding thermal bistabilities which become a severe limitation at high laser powers~\cite{Bhattacharya2007a,ni2012enhancement}. With a $Q_\mathrm{m}$$\sim$$10^8$, our dissipation rates are only matched in other silicon nitride membranes with the use of cryogenic cooling near 14~mK, with an improvement of 2 orders of magnitude in mechanical quality factor from room temperature~\cite{yuan2015silicon}. This allows one to speculate that coupling our new generation of resonators to such low temperature baths could yield $Q_\mathrm{m}$$\sim$1~billion.

Our devices have the potential to allow for fundamental tests of quantum physics by generating massive, nonclassical states of a mechanical oscillator, for example in space, where complicated cryogenic setups are not feasible~\cite{pilan2015performance}. In addition, thanks to the ultralow mechanical dissipation, it is possible to push boundaries of applications in ultrasensitive (e.g.\ force) detection~\cite{Rugar2004,Tao2014,Vinante2016}, as has recently been demonstrated~\cite{Reinhardt2016}. For the devices used here, we calculate a force sensitivity of about 10~aN/Hz$^{1/2}$, which, together with Ref.~\cite{Reinhardt2016} and to our best knowledge, is the highest to date at room temperature. Achieving this level of dissipation in an on-chip design heralds a realistic building block towards optically linked silicon-based quantum networks~\cite{Kimble2008} operating at room temperature.

\def\urlprefix{}

\textbf{Acknowledgments:} We would like to thank M.\ Aspelmeyer, L.\ Bavdaz, A.\ G.\ Krause, I.\ Marinkovi\'{c}, and G.\ A.\ Steele for fruitful discussions and help with early measurements. We also acknowledge valuable support from the Kavli Nanolab Delft, in particular, from C.\ de Boer and M.\ Zuiddam. This project was supported by the European Research Council (ERC StG Strong-Q, grant agreement 676842) and by the Netherlands Organisation for Scientific Research (NWO/OCW), as part of the Frontiers of Nanoscience program.


\begin{thebibliography}{54}
	\expandafter\ifx\csname natexlab\endcsname\relax\def\natexlab#1{#1}\fi
	\expandafter\ifx\csname bibnamefont\endcsname\relax
	\def\bibnamefont#1{#1}\fi
	\expandafter\ifx\csname bibfnamefont\endcsname\relax
	\def\bibfnamefont#1{#1}\fi
	\expandafter\ifx\csname citenamefont\endcsname\relax
	\def\citenamefont#1{#1}\fi
	\expandafter\ifx\csname url\endcsname\relax
	\def\url#1{\texttt{#1}}\fi
	\expandafter\ifx\csname urlprefix\endcsname\relax\def\urlprefix{URL }\fi
	\providecommand{\bibinfo}[2]{#2}
	\providecommand{\eprint}[2][]{\url{#2}}
	
	\bibitem[{\citenamefont{Aspelmeyer et~al.}(2014)\citenamefont{Aspelmeyer,
			Kippenberg, and Marquardt}}]{Aspelmeyer2014}
	\bibinfo{author}{\bibfnamefont{M.}~\bibnamefont{Aspelmeyer}},
	\bibinfo{author}{\bibfnamefont{T.~J.} \bibnamefont{Kippenberg}},
	\bibnamefont{and}
	\bibinfo{author}{\bibfnamefont{F.}~\bibnamefont{Marquardt}},
	\bibinfo{journal}{Rev.\ Mod.\ Phys.} \textbf{\bibinfo{volume}{86}},
	\bibinfo{pages}{1391} (\bibinfo{year}{2014}).
	
	\bibitem[{\citenamefont{Chan et~al.}(2011)\citenamefont{Chan, Alegre,
			Safavi-Naeini, Hill, Krause, Gr\"oblacher, Aspelmeyer, and
			Painter}}]{Chan2011}
	\bibinfo{author}{\bibfnamefont{J.}~\bibnamefont{Chan}},
	\bibinfo{author}{\bibfnamefont{T.~P.~M.} \bibnamefont{Alegre}},
	\bibinfo{author}{\bibfnamefont{A.~H.} \bibnamefont{Safavi-Naeini}},
	\bibinfo{author}{\bibfnamefont{J.~T.} \bibnamefont{Hill}},
	\bibinfo{author}{\bibfnamefont{A.}~\bibnamefont{Krause}},
	\bibinfo{author}{\bibfnamefont{S.}~\bibnamefont{Gr\"oblacher}},
	\bibinfo{author}{\bibfnamefont{M.}~\bibnamefont{Aspelmeyer}},
	\bibnamefont{and} \bibinfo{author}{\bibfnamefont{O.}~\bibnamefont{Painter}},
	\bibinfo{journal}{Nature} \textbf{\bibinfo{volume}{478}}, \bibinfo{pages}{89}
	(\bibinfo{year}{2011}).
	
	\bibitem[{\citenamefont{Teufel et~al.}(2011)\citenamefont{Teufel, Donner, Li,
			Harlow, Allman, Cicak, Sirois, Whittaker, Lehnert, and
			Simmonds}}]{Teufel2011b}
	\bibinfo{author}{\bibfnamefont{J.~D.} \bibnamefont{Teufel}},
	\bibinfo{author}{\bibfnamefont{T.}~\bibnamefont{Donner}},
	\bibinfo{author}{\bibfnamefont{D.}~\bibnamefont{Li}},
	\bibinfo{author}{\bibfnamefont{J.~W.} \bibnamefont{Harlow}},
	\bibinfo{author}{\bibfnamefont{M.~S.} \bibnamefont{Allman}},
	\bibinfo{author}{\bibfnamefont{K.}~\bibnamefont{Cicak}},
	\bibinfo{author}{\bibfnamefont{A.~J.} \bibnamefont{Sirois}},
	\bibinfo{author}{\bibfnamefont{J.~D.} \bibnamefont{Whittaker}},
	\bibinfo{author}{\bibfnamefont{K.~W.} \bibnamefont{Lehnert}},
	\bibnamefont{and} \bibinfo{author}{\bibfnamefont{R.~W.}
		\bibnamefont{Simmonds}}, \bibinfo{journal}{Nature}
	\textbf{\bibinfo{volume}{475}}, \bibinfo{pages}{359} (\bibinfo{year}{2011}).
	
	\bibitem[{\citenamefont{O'Connell et~al.}(2010)\citenamefont{O'Connell,
			Hofheinz, Ansmann, Bialczak, Lenander, Lucero, Neeley, Sank, Wang, Weides
			et~al.}}]{OConnell2010}
	\bibinfo{author}{\bibfnamefont{A.~D.} \bibnamefont{O'Connell}},
	\bibinfo{author}{\bibfnamefont{M.}~\bibnamefont{Hofheinz}},
	\bibinfo{author}{\bibfnamefont{M.}~\bibnamefont{Ansmann}},
	\bibinfo{author}{\bibfnamefont{R.~C.} \bibnamefont{Bialczak}},
	\bibinfo{author}{\bibfnamefont{M.}~\bibnamefont{Lenander}},
	\bibinfo{author}{\bibfnamefont{E.}~\bibnamefont{Lucero}},
	\bibinfo{author}{\bibfnamefont{M.}~\bibnamefont{Neeley}},
	\bibinfo{author}{\bibfnamefont{D.}~\bibnamefont{Sank}},
	\bibinfo{author}{\bibfnamefont{H.}~\bibnamefont{Wang}},
	\bibinfo{author}{\bibfnamefont{M.}~\bibnamefont{Weides}},
	\bibnamefont{et~al.}, \bibinfo{journal}{Nature}
	\textbf{\bibinfo{volume}{464}}, \bibinfo{pages}{697} (\bibinfo{year}{2010}).
	
	\bibitem[{\citenamefont{Riedinger et~al.}(2016)\citenamefont{Riedinger, Hong,
			Norte, Slater, Shang, Krause, Anant, Aspelmeyer, and
			Gr\"oblacher}}]{Riedinger2016}
	\bibinfo{author}{\bibfnamefont{R.}~\bibnamefont{Riedinger}},
	\bibinfo{author}{\bibfnamefont{S.}~\bibnamefont{Hong}},
	\bibinfo{author}{\bibfnamefont{R.~A.} \bibnamefont{Norte}},
	\bibinfo{author}{\bibfnamefont{J.~A.} \bibnamefont{Slater}},
	\bibinfo{author}{\bibfnamefont{J.}~\bibnamefont{Shang}},
	\bibinfo{author}{\bibfnamefont{A.~G.} \bibnamefont{Krause}},
	\bibinfo{author}{\bibfnamefont{V.}~\bibnamefont{Anant}},
	\bibinfo{author}{\bibfnamefont{M.}~\bibnamefont{Aspelmeyer}},
	\bibnamefont{and}
	\bibinfo{author}{\bibfnamefont{S.}~\bibnamefont{Gr\"oblacher}},
	\bibinfo{journal}{Nature} \textbf{\bibinfo{volume}{530}},
	\bibinfo{pages}{313} (\bibinfo{year}{2016}).
	
	\bibitem[{\citenamefont{Palomaki et~al.}(2013)\citenamefont{Palomaki, Teufel,
			Simmonds, and Lehnert}}]{Palomaki2013}
	\bibinfo{author}{\bibfnamefont{T.}~\bibnamefont{Palomaki}},
	\bibinfo{author}{\bibfnamefont{J.}~\bibnamefont{Teufel}},
	\bibinfo{author}{\bibfnamefont{R.}~\bibnamefont{Simmonds}}, \bibnamefont{and}
	\bibinfo{author}{\bibfnamefont{K.}~\bibnamefont{Lehnert}},
	\bibinfo{journal}{Science} \textbf{\bibinfo{volume}{342}},
	\bibinfo{pages}{710} (\bibinfo{year}{2013}).
	
	\bibitem[{\citenamefont{Safavi-Naeini
			et~al.}(2013{\natexlab{a}})\citenamefont{Safavi-Naeini, Gr\"oblacher, Hill,
			Chan, Aspelmeyer, and Painter}}]{Safavi-Naeini2013b}
	\bibinfo{author}{\bibfnamefont{A.~H.} \bibnamefont{Safavi-Naeini}},
	\bibinfo{author}{\bibfnamefont{S.}~\bibnamefont{Gr\"oblacher}},
	\bibinfo{author}{\bibfnamefont{J.~T.} \bibnamefont{Hill}},
	\bibinfo{author}{\bibfnamefont{J.}~\bibnamefont{Chan}},
	\bibinfo{author}{\bibfnamefont{M.}~\bibnamefont{Aspelmeyer}},
	\bibnamefont{and} \bibinfo{author}{\bibfnamefont{O.}~\bibnamefont{Painter}},
	\bibinfo{journal}{Nature} \textbf{\bibinfo{volume}{500}},
	\bibinfo{pages}{185} (\bibinfo{year}{2013}{\natexlab{a}}).
	
	\bibitem[{\citenamefont{Purdy et~al.}(2013)\citenamefont{Purdy, Yu, Peterson,
			Kampel, and Regal}}]{Purdy2013b}
	\bibinfo{author}{\bibfnamefont{T.~P.} \bibnamefont{Purdy}},
	\bibinfo{author}{\bibfnamefont{P.-L.} \bibnamefont{Yu}},
	\bibinfo{author}{\bibfnamefont{R.~W.} \bibnamefont{Peterson}},
	\bibinfo{author}{\bibfnamefont{N.~S.} \bibnamefont{Kampel}},
	\bibnamefont{and} \bibinfo{author}{\bibfnamefont{C.~A.} \bibnamefont{Regal}},
	\bibinfo{journal}{Phys.\ Rev.\ X} \textbf{\bibinfo{volume}{3}},
	\bibinfo{pages}{031012} (\bibinfo{year}{2013}).
	
	\bibitem[{\citenamefont{Wollman et~al.}(2015)\citenamefont{Wollman, Lei,
			Weinstein, Suh, Kronwald, Marquardt, Clerk, and Schwab}}]{wollman2015quantum}
	\bibinfo{author}{\bibfnamefont{E.~E.} \bibnamefont{Wollman}},
	\bibinfo{author}{\bibfnamefont{C.}~\bibnamefont{Lei}},
	\bibinfo{author}{\bibfnamefont{A.}~\bibnamefont{Weinstein}},
	\bibinfo{author}{\bibfnamefont{J.}~\bibnamefont{Suh}},
	\bibinfo{author}{\bibfnamefont{A.}~\bibnamefont{Kronwald}},
	\bibinfo{author}{\bibfnamefont{F.}~\bibnamefont{Marquardt}},
	\bibinfo{author}{\bibfnamefont{A.}~\bibnamefont{Clerk}}, \bibnamefont{and}
	\bibinfo{author}{\bibfnamefont{K.}~\bibnamefont{Schwab}},
	\bibinfo{journal}{Science} \textbf{\bibinfo{volume}{349}},
	\bibinfo{pages}{952} (\bibinfo{year}{2015}).
	
	\bibitem[{\citenamefont{Pirkkalainen et~al.}(2015)\citenamefont{Pirkkalainen,
			Damsk\"agg, Brandt, Massel, and Sillanp\"a\"a}}]{Pirkkalainen2015}
	\bibinfo{author}{\bibfnamefont{J.-M.} \bibnamefont{Pirkkalainen}},
	\bibinfo{author}{\bibfnamefont{E.}~\bibnamefont{Damsk\"agg}},
	\bibinfo{author}{\bibfnamefont{M.}~\bibnamefont{Brandt}},
	\bibinfo{author}{\bibfnamefont{F.}~\bibnamefont{Massel}}, \bibnamefont{and}
	\bibinfo{author}{\bibfnamefont{M.~A.} \bibnamefont{Sillanp\"a\"a}},
	\bibinfo{journal}{Phys.\ Rev.\ Lett.} \textbf{\bibinfo{volume}{115}},
	\bibinfo{pages}{243601} (\bibinfo{year}{2015}).
	
	\bibitem[{\citenamefont{Lecocq et~al.}(2015)\citenamefont{Lecocq, Clark,
			Simmonds, Aumentado, and Teufel}}]{Lecocq2015}
	\bibinfo{author}{\bibfnamefont{F.}~\bibnamefont{Lecocq}},
	\bibinfo{author}{\bibfnamefont{J.~B.} \bibnamefont{Clark}},
	\bibinfo{author}{\bibfnamefont{R.~W.} \bibnamefont{Simmonds}},
	\bibinfo{author}{\bibfnamefont{J.}~\bibnamefont{Aumentado}},
	\bibnamefont{and} \bibinfo{author}{\bibfnamefont{J.~D.}
		\bibnamefont{Teufel}}, \bibinfo{journal}{Phys.\ Rev.\ X}
	\textbf{\bibinfo{volume}{5}}, \bibinfo{pages}{041037} (\bibinfo{year}{2015}).
	
	\bibitem[{\citenamefont{Treutlein et~al.}(2007)\citenamefont{Treutlein, Hunger,
			Camerer, H\"ansch, and Reichel}}]{Treutlein2007}
	\bibinfo{author}{\bibfnamefont{P.}~\bibnamefont{Treutlein}},
	\bibinfo{author}{\bibfnamefont{D.}~\bibnamefont{Hunger}},
	\bibinfo{author}{\bibfnamefont{S.}~\bibnamefont{Camerer}},
	\bibinfo{author}{\bibfnamefont{T.~W.} \bibnamefont{H\"ansch}},
	\bibnamefont{and} \bibinfo{author}{\bibfnamefont{J.}~\bibnamefont{Reichel}},
	\bibinfo{journal}{Phys.\ Rev.\ Lett.} \textbf{\bibinfo{volume}{99}},
	\bibinfo{pages}{140403} (\bibinfo{year}{2007}).
	
	\bibitem[{\citenamefont{Hammerer et~al.}(2009)\citenamefont{Hammerer,
			Wallquist, Genes, Ludwig, Marquardt, Treutlein, Zoller, Ye, and
			Kimble}}]{Hammerer2009b}
	\bibinfo{author}{\bibfnamefont{K.}~\bibnamefont{Hammerer}},
	\bibinfo{author}{\bibfnamefont{M.}~\bibnamefont{Wallquist}},
	\bibinfo{author}{\bibfnamefont{C.}~\bibnamefont{Genes}},
	\bibinfo{author}{\bibfnamefont{M.}~\bibnamefont{Ludwig}},
	\bibinfo{author}{\bibfnamefont{F.}~\bibnamefont{Marquardt}},
	\bibinfo{author}{\bibfnamefont{P.}~\bibnamefont{Treutlein}},
	\bibinfo{author}{\bibfnamefont{P.}~\bibnamefont{Zoller}},
	\bibinfo{author}{\bibfnamefont{J.}~\bibnamefont{Ye}}, \bibnamefont{and}
	\bibinfo{author}{\bibfnamefont{H.~J.} \bibnamefont{Kimble}},
	\bibinfo{journal}{Phys.\ Rev.\ Lett.} \textbf{\bibinfo{volume}{103}},
	\bibinfo{pages}{063005} (\bibinfo{year}{2009}).
	
	\bibitem[{\citenamefont{Krause et~al.}(2012)\citenamefont{Krause, Winger,
			Blasius, Lin, and Painter}}]{krause2012high}
	\bibinfo{author}{\bibfnamefont{A.~G.} \bibnamefont{Krause}},
	\bibinfo{author}{\bibfnamefont{M.}~\bibnamefont{Winger}},
	\bibinfo{author}{\bibfnamefont{T.~D.} \bibnamefont{Blasius}},
	\bibinfo{author}{\bibfnamefont{Q.}~\bibnamefont{Lin}}, \bibnamefont{and}
	\bibinfo{author}{\bibfnamefont{O.}~\bibnamefont{Painter}},
	\bibinfo{journal}{Nature Photon.} \textbf{\bibinfo{volume}{6}},
	\bibinfo{pages}{768} (\bibinfo{year}{2012}).
	
	\bibitem[{\citenamefont{J{\"o}ckel et~al.}(2014)\citenamefont{J{\"o}ckel,
			Faber, Kampschulte, Korppi, Rakher, and Treutlein}}]{jockel2014sympathetic}
	\bibinfo{author}{\bibfnamefont{A.}~\bibnamefont{J{\"o}ckel}},
	\bibinfo{author}{\bibfnamefont{A.}~\bibnamefont{Faber}},
	\bibinfo{author}{\bibfnamefont{T.}~\bibnamefont{Kampschulte}},
	\bibinfo{author}{\bibfnamefont{M.}~\bibnamefont{Korppi}},
	\bibinfo{author}{\bibfnamefont{M.~T.} \bibnamefont{Rakher}},
	\bibnamefont{and}
	\bibinfo{author}{\bibfnamefont{P.}~\bibnamefont{Treutlein}},
	\bibinfo{journal}{Nature Nanotech.} \textbf{\bibinfo{volume}{10}},
	\bibinfo{pages}{55} (\bibinfo{year}{2014}).
	
	\bibitem[{\citenamefont{Stamper-Kurn}(2014)}]{Stamper-Kurn2014}
	\bibinfo{author}{\bibfnamefont{D.~M.} \bibnamefont{Stamper-Kurn}},
	\emph{\bibinfo{title}{Cavity Optomechanics}} (\bibinfo{publisher}{Springer},
	\bibinfo{year}{2014}), chap. \bibinfo{chapter}{Cavity Optomechanics with Cold
		Atoms}, pp. \bibinfo{pages}{283--325}.
	
	\bibitem[{\citenamefont{Yuan et~al.}(2015{\natexlab{a}})\citenamefont{Yuan,
			Singh, Blanter, and Steele}}]{yuan2015large}
	\bibinfo{author}{\bibfnamefont{M.}~\bibnamefont{Yuan}},
	\bibinfo{author}{\bibfnamefont{V.}~\bibnamefont{Singh}},
	\bibinfo{author}{\bibfnamefont{Y.~M.} \bibnamefont{Blanter}},
	\bibnamefont{and} \bibinfo{author}{\bibfnamefont{G.~A.}
		\bibnamefont{Steele}}, \bibinfo{journal}{Nature Commun.}
	\textbf{\bibinfo{volume}{6}}, \bibinfo{pages}{8491}
	(\bibinfo{year}{2015}{\natexlab{a}}).
	
	\bibitem[{\citenamefont{Andrews et~al.}(2014)\citenamefont{Andrews, Peterson,
			Purdy, Cicak, Simmonds, Regal, and Lehnert}}]{andrews2014bidirectional}
	\bibinfo{author}{\bibfnamefont{R.}~\bibnamefont{Andrews}},
	\bibinfo{author}{\bibfnamefont{R.}~\bibnamefont{Peterson}},
	\bibinfo{author}{\bibfnamefont{T.}~\bibnamefont{Purdy}},
	\bibinfo{author}{\bibfnamefont{K.}~\bibnamefont{Cicak}},
	\bibinfo{author}{\bibfnamefont{R.}~\bibnamefont{Simmonds}},
	\bibinfo{author}{\bibfnamefont{C.}~\bibnamefont{Regal}}, \bibnamefont{and}
	\bibinfo{author}{\bibfnamefont{K.}~\bibnamefont{Lehnert}},
	\bibinfo{journal}{Nature Phys.} \textbf{\bibinfo{volume}{10}},
	\bibinfo{pages}{321} (\bibinfo{year}{2014}).
	
	\bibitem[{\citenamefont{Thompson et~al.}(2008)\citenamefont{Thompson, Zwickl,
			Jayich, Marquardt, Girvin, and Harris}}]{thompson2008strong}
	\bibinfo{author}{\bibfnamefont{J.}~\bibnamefont{Thompson}},
	\bibinfo{author}{\bibfnamefont{B.}~\bibnamefont{Zwickl}},
	\bibinfo{author}{\bibfnamefont{A.}~\bibnamefont{Jayich}},
	\bibinfo{author}{\bibfnamefont{F.}~\bibnamefont{Marquardt}},
	\bibinfo{author}{\bibfnamefont{S.}~\bibnamefont{Girvin}}, \bibnamefont{and}
	\bibinfo{author}{\bibfnamefont{J.}~\bibnamefont{Harris}},
	\bibinfo{journal}{Nature} \textbf{\bibinfo{volume}{452}}, \bibinfo{pages}{72}
	(\bibinfo{year}{2008}).
	
	\bibitem[{\citenamefont{Marquardt et~al.}(2007)\citenamefont{Marquardt, Chen,
			Clerk, and Girvin}}]{Marquardt2007}
	\bibinfo{author}{\bibfnamefont{F.}~\bibnamefont{Marquardt}},
	\bibinfo{author}{\bibfnamefont{J.~P.} \bibnamefont{Chen}},
	\bibinfo{author}{\bibfnamefont{A.~A.} \bibnamefont{Clerk}}, \bibnamefont{and}
	\bibinfo{author}{\bibfnamefont{S.~M.} \bibnamefont{Girvin}},
	\bibinfo{journal}{Phys.\ Rev.\ Lett.} \textbf{\bibinfo{volume}{99}},
	\bibinfo{pages}{093902} (\bibinfo{year}{2007}).
	
	\bibitem[{\citenamefont{Wilson et~al.}(2009)\citenamefont{Wilson, Regal, Papp,
			and Kimble}}]{Wilson2009}
	\bibinfo{author}{\bibfnamefont{D.~J.} \bibnamefont{Wilson}},
	\bibinfo{author}{\bibfnamefont{C.~A.} \bibnamefont{Regal}},
	\bibinfo{author}{\bibfnamefont{S.~B.} \bibnamefont{Papp}}, \bibnamefont{and}
	\bibinfo{author}{\bibfnamefont{H.~J.} \bibnamefont{Kimble}},
	\bibinfo{journal}{Phys.\ Rev.\ Lett.} \textbf{\bibinfo{volume}{103}},
	\bibinfo{pages}{207204} (\bibinfo{year}{2009}).
	
	\bibitem[{\citenamefont{Chakram et~al.}(2014)\citenamefont{Chakram, Patil,
			Chang, and Vengalattore}}]{Chakram2014}
	\bibinfo{author}{\bibfnamefont{S.}~\bibnamefont{Chakram}},
	\bibinfo{author}{\bibfnamefont{Y.~S.} \bibnamefont{Patil}},
	\bibinfo{author}{\bibfnamefont{L.}~\bibnamefont{Chang}}, \bibnamefont{and}
	\bibinfo{author}{\bibfnamefont{M.}~\bibnamefont{Vengalattore}},
	\bibinfo{journal}{Phys. Rev. Lett.} \textbf{\bibinfo{volume}{112}},
	\bibinfo{pages}{127201} (\bibinfo{year}{2014}).
	
	\bibitem[{\citenamefont{Safavi-Naeini
			et~al.}(2013{\natexlab{b}})\citenamefont{Safavi-Naeini, Chan, Hill,
			Gr\"oblacher, Miao, Chen, Aspelmeyer, and Painter}}]{Safavi-Naeini2013a}
	\bibinfo{author}{\bibfnamefont{A.~H.} \bibnamefont{Safavi-Naeini}},
	\bibinfo{author}{\bibfnamefont{J.}~\bibnamefont{Chan}},
	\bibinfo{author}{\bibfnamefont{J.~T.} \bibnamefont{Hill}},
	\bibinfo{author}{\bibfnamefont{S.}~\bibnamefont{Gr\"oblacher}},
	\bibinfo{author}{\bibfnamefont{H.}~\bibnamefont{Miao}},
	\bibinfo{author}{\bibfnamefont{Y.}~\bibnamefont{Chen}},
	\bibinfo{author}{\bibfnamefont{M.}~\bibnamefont{Aspelmeyer}},
	\bibnamefont{and} \bibinfo{author}{\bibfnamefont{O.}~\bibnamefont{Painter}},
	\bibinfo{journal}{New J.\ Phys.} \textbf{\bibinfo{volume}{15}},
	\bibinfo{pages}{035007} (\bibinfo{year}{2013}{\natexlab{b}}).
	
	\bibitem[{\citenamefont{Harry et~al.}(2002)\citenamefont{Harry, Gretarsson,
			Saulson, Kittelberger, Penn, Startin, Rowan, Fejer, Crooks, Cagnoli
			et~al.}}]{Harry2002}
	\bibinfo{author}{\bibfnamefont{G.~M.} \bibnamefont{Harry}},
	\bibinfo{author}{\bibfnamefont{A.~M.} \bibnamefont{Gretarsson}},
	\bibinfo{author}{\bibfnamefont{P.~R.} \bibnamefont{Saulson}},
	\bibinfo{author}{\bibfnamefont{S.~E.} \bibnamefont{Kittelberger}},
	\bibinfo{author}{\bibfnamefont{S.~D.} \bibnamefont{Penn}},
	\bibinfo{author}{\bibfnamefont{W.~J.} \bibnamefont{Startin}},
	\bibinfo{author}{\bibfnamefont{S.}~\bibnamefont{Rowan}},
	\bibinfo{author}{\bibfnamefont{M.~M.} \bibnamefont{Fejer}},
	\bibinfo{author}{\bibfnamefont{D.~R.~M.} \bibnamefont{Crooks}},
	\bibinfo{author}{\bibfnamefont{G.}~\bibnamefont{Cagnoli}},
	\bibnamefont{et~al.}, \bibinfo{journal}{Class.\ Quantum Grav.}
	\textbf{\bibinfo{volume}{19}}, \bibinfo{pages}{897} (\bibinfo{year}{2002}).
	
	\bibitem[{\citenamefont{Gr\"{o}blacher
			et~al.}(2009)\citenamefont{Gr\"{o}blacher, Hertzberg, Vanner, Gigan, Schwab,
			and Aspelmeyer}}]{Groeblacher2009a}
	\bibinfo{author}{\bibfnamefont{S.}~\bibnamefont{Gr\"{o}blacher}},
	\bibinfo{author}{\bibfnamefont{J.~B.} \bibnamefont{Hertzberg}},
	\bibinfo{author}{\bibfnamefont{M.~R.} \bibnamefont{Vanner}},
	\bibinfo{author}{\bibfnamefont{S.}~\bibnamefont{Gigan}},
	\bibinfo{author}{\bibfnamefont{K.~C.} \bibnamefont{Schwab}},
	\bibnamefont{and}
	\bibinfo{author}{\bibfnamefont{M.}~\bibnamefont{Aspelmeyer}},
	\bibinfo{journal}{Nature Phys.} \textbf{\bibinfo{volume}{5}},
	\bibinfo{pages}{485} (\bibinfo{year}{2009}).
	
	\bibitem[{\citenamefont{Kleckner et~al.}(2011)\citenamefont{Kleckner, Pepper,
			Jeffrey, Sonin, Thon, and Bouwmeester}}]{kleckner2011optomechanical}
	\bibinfo{author}{\bibfnamefont{D.}~\bibnamefont{Kleckner}},
	\bibinfo{author}{\bibfnamefont{B.}~\bibnamefont{Pepper}},
	\bibinfo{author}{\bibfnamefont{E.}~\bibnamefont{Jeffrey}},
	\bibinfo{author}{\bibfnamefont{P.}~\bibnamefont{Sonin}},
	\bibinfo{author}{\bibfnamefont{S.~M.} \bibnamefont{Thon}}, \bibnamefont{and}
	\bibinfo{author}{\bibfnamefont{D.}~\bibnamefont{Bouwmeester}},
	\bibinfo{journal}{Opt.\ Express} \textbf{\bibinfo{volume}{19}},
	\bibinfo{pages}{19708} (\bibinfo{year}{2011}).
	
	\bibitem[{\citenamefont{Chang et~al.}(2010)\citenamefont{Chang, Regal, Papp,
			Wilson, Ye, Painter, Kimble, and Zoller}}]{Chang2010}
	\bibinfo{author}{\bibfnamefont{D.~E.} \bibnamefont{Chang}},
	\bibinfo{author}{\bibfnamefont{C.~A.} \bibnamefont{Regal}},
	\bibinfo{author}{\bibfnamefont{S.~B.} \bibnamefont{Papp}},
	\bibinfo{author}{\bibfnamefont{D.~J.} \bibnamefont{Wilson}},
	\bibinfo{author}{\bibfnamefont{J.}~\bibnamefont{Ye}},
	\bibinfo{author}{\bibfnamefont{O.}~\bibnamefont{Painter}},
	\bibinfo{author}{\bibfnamefont{H.~J.} \bibnamefont{Kimble}},
	\bibnamefont{and} \bibinfo{author}{\bibfnamefont{P.}~\bibnamefont{Zoller}},
	\bibinfo{journal}{Proc.\ Natl.\ Acad.\ Sci.} \textbf{\bibinfo{volume}{107}},
	\bibinfo{pages}{1005} (\bibinfo{year}{2010}).
	
	\bibitem[{\citenamefont{Gieseler et~al.}(2013)\citenamefont{Gieseler, Novotny,
			and Quidant}}]{gieseler2013thermal}
	\bibinfo{author}{\bibfnamefont{J.}~\bibnamefont{Gieseler}},
	\bibinfo{author}{\bibfnamefont{L.}~\bibnamefont{Novotny}}, \bibnamefont{and}
	\bibinfo{author}{\bibfnamefont{R.}~\bibnamefont{Quidant}},
	\bibinfo{journal}{Nature Phys.} \textbf{\bibinfo{volume}{9}},
	\bibinfo{pages}{806} (\bibinfo{year}{2013}).
	
	\bibitem[{\citenamefont{Zhang et~al.}(2015)\citenamefont{Zhang, Ti, Davanço,
			Ren, Aksyuk, Liu, and Srinivasan}}]{Zhang2015}
	\bibinfo{author}{\bibfnamefont{R.}~\bibnamefont{Zhang}},
	\bibinfo{author}{\bibfnamefont{C.}~\bibnamefont{Ti}},
	\bibinfo{author}{\bibfnamefont{M.~I.} \bibnamefont{Davanço}},
	\bibinfo{author}{\bibfnamefont{Y.}~\bibnamefont{Ren}},
	\bibinfo{author}{\bibfnamefont{V.}~\bibnamefont{Aksyuk}},
	\bibinfo{author}{\bibfnamefont{Y.}~\bibnamefont{Liu}}, \bibnamefont{and}
	\bibinfo{author}{\bibfnamefont{K.}~\bibnamefont{Srinivasan}},
	\bibinfo{journal}{Appl.\ Phys.\ Lett.} \textbf{\bibinfo{volume}{107}},
	\bibinfo{pages}{131110} (\bibinfo{year}{2015}).
	
	\bibitem[{\citenamefont{Nunnenkamp et~al.}(2010)\citenamefont{Nunnenkamp,
			B{\o}rkje, Harris, and Girvin}}]{nunnenkamp2010cooling}
	\bibinfo{author}{\bibfnamefont{A.}~\bibnamefont{Nunnenkamp}},
	\bibinfo{author}{\bibfnamefont{K.}~\bibnamefont{B{\o}rkje}},
	\bibinfo{author}{\bibfnamefont{J.}~\bibnamefont{Harris}}, \bibnamefont{and}
	\bibinfo{author}{\bibfnamefont{S.}~\bibnamefont{Girvin}},
	\bibinfo{journal}{Phys.\ Rev.\ A} \textbf{\bibinfo{volume}{82}},
	\bibinfo{pages}{021806} (\bibinfo{year}{2010}).
	
	\bibitem[{\citenamefont{Deng et~al.}(2012)\citenamefont{Deng, Li, Gao, and
			Wu}}]{deng2012performance}
	\bibinfo{author}{\bibfnamefont{Z.}~\bibnamefont{Deng}},
	\bibinfo{author}{\bibfnamefont{Y.}~\bibnamefont{Li}},
	\bibinfo{author}{\bibfnamefont{M.}~\bibnamefont{Gao}}, \bibnamefont{and}
	\bibinfo{author}{\bibfnamefont{C.}~\bibnamefont{Wu}},
	\bibinfo{journal}{Phys.\ Rev.\ A} \textbf{\bibinfo{volume}{85}},
	\bibinfo{pages}{025804} (\bibinfo{year}{2012}).
	
	\bibitem[{\citenamefont{Usami et~al.}(2012)\citenamefont{Usami, Naesby, Bagci,
			Nielsen, Liu, Stobbe, Lodahl, and Polzik}}]{usami2012optical}
	\bibinfo{author}{\bibfnamefont{K.}~\bibnamefont{Usami}},
	\bibinfo{author}{\bibfnamefont{A.}~\bibnamefont{Naesby}},
	\bibinfo{author}{\bibfnamefont{T.}~\bibnamefont{Bagci}},
	\bibinfo{author}{\bibfnamefont{B.~M.} \bibnamefont{Nielsen}},
	\bibinfo{author}{\bibfnamefont{J.}~\bibnamefont{Liu}},
	\bibinfo{author}{\bibfnamefont{S.}~\bibnamefont{Stobbe}},
	\bibinfo{author}{\bibfnamefont{P.}~\bibnamefont{Lodahl}}, \bibnamefont{and}
	\bibinfo{author}{\bibfnamefont{E.~S.} \bibnamefont{Polzik}},
	\bibinfo{journal}{Nature Phys.} \textbf{\bibinfo{volume}{8}},
	\bibinfo{pages}{168} (\bibinfo{year}{2012}).
	
	\bibitem[{\citenamefont{Jayich et~al.}(2008)\citenamefont{Jayich, Sankey,
			Zwickl, Yang, Thompson, Girvin, Clerk, Marquardt, and
			Harris}}]{jayich2008dispersive}
	\bibinfo{author}{\bibfnamefont{A.}~\bibnamefont{Jayich}},
	\bibinfo{author}{\bibfnamefont{J.}~\bibnamefont{Sankey}},
	\bibinfo{author}{\bibfnamefont{B.}~\bibnamefont{Zwickl}},
	\bibinfo{author}{\bibfnamefont{C.}~\bibnamefont{Yang}},
	\bibinfo{author}{\bibfnamefont{J.}~\bibnamefont{Thompson}},
	\bibinfo{author}{\bibfnamefont{S.}~\bibnamefont{Girvin}},
	\bibinfo{author}{\bibfnamefont{A.}~\bibnamefont{Clerk}},
	\bibinfo{author}{\bibfnamefont{F.}~\bibnamefont{Marquardt}},
	\bibnamefont{and} \bibinfo{author}{\bibfnamefont{J.}~\bibnamefont{Harris}},
	\bibinfo{journal}{New J.\ Phys.} \textbf{\bibinfo{volume}{10}},
	\bibinfo{pages}{095008} (\bibinfo{year}{2008}).
	
	\bibitem[{\citenamefont{Gigan et~al.}(2006)\citenamefont{Gigan, B\"ohm,
			Paternostro, Blaser, Langer, Hertzberg, Schwab, B\"auerle, Aspelmeyer, and
			Zeilinger}}]{Gigan2006}
	\bibinfo{author}{\bibfnamefont{S.}~\bibnamefont{Gigan}},
	\bibinfo{author}{\bibfnamefont{H.~R.} \bibnamefont{B\"ohm}},
	\bibinfo{author}{\bibfnamefont{M.}~\bibnamefont{Paternostro}},
	\bibinfo{author}{\bibfnamefont{F.}~\bibnamefont{Blaser}},
	\bibinfo{author}{\bibfnamefont{G.}~\bibnamefont{Langer}},
	\bibinfo{author}{\bibfnamefont{J.~B.} \bibnamefont{Hertzberg}},
	\bibinfo{author}{\bibfnamefont{K.~C.} \bibnamefont{Schwab}},
	\bibinfo{author}{\bibfnamefont{D.}~\bibnamefont{B\"auerle}},
	\bibinfo{author}{\bibfnamefont{M.}~\bibnamefont{Aspelmeyer}},
	\bibnamefont{and}
	\bibinfo{author}{\bibfnamefont{A.}~\bibnamefont{Zeilinger}},
	\bibinfo{journal}{Nature} \textbf{\bibinfo{volume}{444}}, \bibinfo{pages}{67}
	(\bibinfo{year}{2006}).
	
	\bibitem[{\citenamefont{Arcizet et~al.}(2006)\citenamefont{Arcizet, Cohadon,
			Briant, Pinard, and Heidmann}}]{Arcizet2006b}
	\bibinfo{author}{\bibfnamefont{O.}~\bibnamefont{Arcizet}},
	\bibinfo{author}{\bibfnamefont{P.-F.} \bibnamefont{Cohadon}},
	\bibinfo{author}{\bibfnamefont{T.}~\bibnamefont{Briant}},
	\bibinfo{author}{\bibfnamefont{M.}~\bibnamefont{Pinard}}, \bibnamefont{and}
	\bibinfo{author}{\bibfnamefont{A.}~\bibnamefont{Heidmann}},
	\bibinfo{journal}{Nature} \textbf{\bibinfo{volume}{444}}, \bibinfo{pages}{71}
	(\bibinfo{year}{2006}).
	
	\bibitem[{\citenamefont{Schliesser et~al.}(2006)\citenamefont{Schliesser,
			Del'Haye, Nooshi, Vahala, and Kippenberg}}]{Schliesser2006}
	\bibinfo{author}{\bibfnamefont{A.}~\bibnamefont{Schliesser}},
	\bibinfo{author}{\bibfnamefont{P.}~\bibnamefont{Del'Haye}},
	\bibinfo{author}{\bibfnamefont{N.}~\bibnamefont{Nooshi}},
	\bibinfo{author}{\bibfnamefont{K.~J.} \bibnamefont{Vahala}},
	\bibnamefont{and} \bibinfo{author}{\bibfnamefont{T.~J.}
		\bibnamefont{Kippenberg}}, \bibinfo{journal}{Phys.\ Rev.\ Lett.}
	\textbf{\bibinfo{volume}{97}}, \bibinfo{pages}{243905}
	(\bibinfo{year}{2006}).
	
	\bibitem[{\citenamefont{Stambaugh et~al.}(2015)\citenamefont{Stambaugh, Xu,
			Kemiktarak, Taylor, and Lawall}}]{Stambaugh2015}
	\bibinfo{author}{\bibfnamefont{C.}~\bibnamefont{Stambaugh}},
	\bibinfo{author}{\bibfnamefont{H.}~\bibnamefont{Xu}},
	\bibinfo{author}{\bibfnamefont{U.}~\bibnamefont{Kemiktarak}},
	\bibinfo{author}{\bibfnamefont{J.}~\bibnamefont{Taylor}}, \bibnamefont{and}
	\bibinfo{author}{\bibfnamefont{J.}~\bibnamefont{Lawall}},
	\bibinfo{journal}{Ann.\ Phys.} \textbf{\bibinfo{volume}{527}},
	\bibinfo{pages}{81} (\bibinfo{year}{2015}).
	
	\bibitem[{\citenamefont{Bui et~al.}(2012)\citenamefont{Bui, Zheng, Hoch, Lee,
			Harris, and Wong}}]{Bui2012}
	\bibinfo{author}{\bibfnamefont{C.~H.} \bibnamefont{Bui}},
	\bibinfo{author}{\bibfnamefont{J.}~\bibnamefont{Zheng}},
	\bibinfo{author}{\bibfnamefont{S.~W.} \bibnamefont{Hoch}},
	\bibinfo{author}{\bibfnamefont{L.~Y.~T.} \bibnamefont{Lee}},
	\bibinfo{author}{\bibfnamefont{J.~G.~E.} \bibnamefont{Harris}},
	\bibnamefont{and} \bibinfo{author}{\bibfnamefont{C.~W.} \bibnamefont{Wong}},
	\bibinfo{journal}{Appl.\ Phys.\ Lett.} \textbf{\bibinfo{volume}{100}},
	\bibinfo{pages}{021110} (\bibinfo{year}{2012}).
	
	\bibitem[{\citenamefont{Makles et~al.}(2015)\citenamefont{Makles, Antoni, Kuhn,
			Del\'eglise, Briant, Cohadon, Braive, Beaudoin, Pinard, Michel
			et~al.}}]{Makles2015}
	\bibinfo{author}{\bibfnamefont{K.}~\bibnamefont{Makles}},
	\bibinfo{author}{\bibfnamefont{T.}~\bibnamefont{Antoni}},
	\bibinfo{author}{\bibfnamefont{A.~G.} \bibnamefont{Kuhn}},
	\bibinfo{author}{\bibfnamefont{S.}~\bibnamefont{Del\'eglise}},
	\bibinfo{author}{\bibfnamefont{T.}~\bibnamefont{Briant}},
	\bibinfo{author}{\bibfnamefont{P.-F.} \bibnamefont{Cohadon}},
	\bibinfo{author}{\bibfnamefont{R.}~\bibnamefont{Braive}},
	\bibinfo{author}{\bibfnamefont{G.}~\bibnamefont{Beaudoin}},
	\bibinfo{author}{\bibfnamefont{L.}~\bibnamefont{Pinard}},
	\bibinfo{author}{\bibfnamefont{C.}~\bibnamefont{Michel}},
	\bibnamefont{et~al.}, \bibinfo{journal}{Optics Lett.}
	\textbf{\bibinfo{volume}{40}}, \bibinfo{pages}{174} (\bibinfo{year}{2015}).
	
	\bibitem[{\citenamefont{Verbridge et~al.}(2008)\citenamefont{Verbridge, Ilic,
			Craighead, and Parpia}}]{verbridge2008size}
	\bibinfo{author}{\bibfnamefont{S.~S.} \bibnamefont{Verbridge}},
	\bibinfo{author}{\bibfnamefont{R.}~\bibnamefont{Ilic}},
	\bibinfo{author}{\bibfnamefont{H.}~\bibnamefont{Craighead}},
	\bibnamefont{and} \bibinfo{author}{\bibfnamefont{J.~M.}
		\bibnamefont{Parpia}}, \bibinfo{journal}{Appl.\ Phys.\ Lett.}
	\textbf{\bibinfo{volume}{93}}, \bibinfo{pages}{013101}
	(\bibinfo{year}{2008}).
	
	\bibitem[{\citenamefont{Schmid et~al.}(2011)\citenamefont{Schmid, Jensen,
			Nielsen, and Boisen}}]{schmid2011damping}
	\bibinfo{author}{\bibfnamefont{S.}~\bibnamefont{Schmid}},
	\bibinfo{author}{\bibfnamefont{K.}~\bibnamefont{Jensen}},
	\bibinfo{author}{\bibfnamefont{K.}~\bibnamefont{Nielsen}}, \bibnamefont{and}
	\bibinfo{author}{\bibfnamefont{A.}~\bibnamefont{Boisen}},
	\bibinfo{journal}{Phys.\ Rev.\ B} \textbf{\bibinfo{volume}{84}},
	\bibinfo{pages}{165307} (\bibinfo{year}{2011}).
	
	\bibitem[{\citenamefont{Darvishian et~al.}(2015)\citenamefont{Darvishian,
			Shiari, He, and Najafi}}]{darvishian2015effect}
	\bibinfo{author}{\bibfnamefont{A.}~\bibnamefont{Darvishian}},
	\bibinfo{author}{\bibfnamefont{B.}~\bibnamefont{Shiari}},
	\bibinfo{author}{\bibfnamefont{G.}~\bibnamefont{He}}, \bibnamefont{and}
	\bibinfo{author}{\bibfnamefont{K.}~\bibnamefont{Najafi}}, in
	\emph{\bibinfo{booktitle}{Inertial Sensors and Systems (ISISS), 2015 IEEE
			International Symposium on}} (\bibinfo{organization}{IEEE},
	\bibinfo{year}{2015}), pp. \bibinfo{pages}{1--4}.
	
	\bibitem[{\citenamefont{Wilson}(2012)}]{wilson2012cavity}
	\bibinfo{author}{\bibfnamefont{D.~J.} \bibnamefont{Wilson}}, Ph.D. thesis,
	\bibinfo{school}{California Institute of Technology} (\bibinfo{year}{2012}).
	
	\bibitem[{\citenamefont{Kaushik et~al.}(2005)\citenamefont{Kaushik, Kahn, and
			Heuer}}]{kaushik2005wafer}
	\bibinfo{author}{\bibfnamefont{A.}~\bibnamefont{Kaushik}},
	\bibinfo{author}{\bibfnamefont{H.}~\bibnamefont{Kahn}}, \bibnamefont{and}
	\bibinfo{author}{\bibfnamefont{A.~H.} \bibnamefont{Heuer}},
	\bibinfo{journal}{J.\ Microelectromech.\ Syst.}
	\textbf{\bibinfo{volume}{14}}, \bibinfo{pages}{359} (\bibinfo{year}{2005}).
	
	\bibitem[{\citenamefont{Norte}(2014)}]{norte2014}
	\bibinfo{author}{\bibfnamefont{R.~A.} \bibnamefont{Norte}}, Ph.D. thesis,
	\bibinfo{school}{California Institute of Technology} (\bibinfo{year}{2014}).
	
	\bibitem[{\citenamefont{Bhattacharya and Meystre}(2007)}]{Bhattacharya2007a}
	\bibinfo{author}{\bibfnamefont{M.}~\bibnamefont{Bhattacharya}}
	\bibnamefont{and} \bibinfo{author}{\bibfnamefont{P.}~\bibnamefont{Meystre}},
	\bibinfo{journal}{Phys.\ Rev.\ Lett.} \textbf{\bibinfo{volume}{99}},
	\bibinfo{pages}{073601} (\bibinfo{year}{2007}).
	
	\bibitem[{\citenamefont{Ni et~al.}(2012)\citenamefont{Ni, Norte, Wilson, Hood,
			Chang, Painter, and Kimble}}]{ni2012enhancement}
	\bibinfo{author}{\bibfnamefont{K.-K.} \bibnamefont{Ni}},
	\bibinfo{author}{\bibfnamefont{R.}~\bibnamefont{Norte}},
	\bibinfo{author}{\bibfnamefont{D.}~\bibnamefont{Wilson}},
	\bibinfo{author}{\bibfnamefont{J.}~\bibnamefont{Hood}},
	\bibinfo{author}{\bibfnamefont{D.}~\bibnamefont{Chang}},
	\bibinfo{author}{\bibfnamefont{O.}~\bibnamefont{Painter}}, \bibnamefont{and}
	\bibinfo{author}{\bibfnamefont{H.}~\bibnamefont{Kimble}},
	\bibinfo{journal}{Phys.\ Rev.\ Lett.} \textbf{\bibinfo{volume}{108}},
	\bibinfo{pages}{214302} (\bibinfo{year}{2012}).
	
	\bibitem[{\citenamefont{Yuan et~al.}(2015{\natexlab{b}})\citenamefont{Yuan,
			Cohen, and Steele}}]{yuan2015silicon}
	\bibinfo{author}{\bibfnamefont{M.}~\bibnamefont{Yuan}},
	\bibinfo{author}{\bibfnamefont{M.~A.} \bibnamefont{Cohen}}, \bibnamefont{and}
	\bibinfo{author}{\bibfnamefont{G.}~\bibnamefont{Steele}},
	\bibinfo{journal}{Appl.\ Phys.\ Lett.} \textbf{\bibinfo{volume}{107}},
	\bibinfo{pages}{263501} (\bibinfo{year}{2015}{\natexlab{b}}).
	
	\bibitem[{\citenamefont{Pilan-Zanoni et~al.}(2015)\citenamefont{Pilan-Zanoni,
			Burkhardt, Johann, Aspelmeyer, Kaltenbaek, and
			Hechenblaikner}}]{pilan2015performance}
	\bibinfo{author}{\bibfnamefont{A.}~\bibnamefont{Pilan-Zanoni}},
	\bibinfo{author}{\bibfnamefont{J.}~\bibnamefont{Burkhardt}},
	\bibinfo{author}{\bibfnamefont{U.}~\bibnamefont{Johann}},
	\bibinfo{author}{\bibfnamefont{M.}~\bibnamefont{Aspelmeyer}},
	\bibinfo{author}{\bibfnamefont{R.}~\bibnamefont{Kaltenbaek}},
	\bibnamefont{and}
	\bibinfo{author}{\bibfnamefont{G.}~\bibnamefont{Hechenblaikner}},
	\bibinfo{journal}{arXiv:1508.01032}  (\bibinfo{year}{2015}).
	
	\bibitem[{\citenamefont{Rugar et~al.}(2004)\citenamefont{Rugar, Budakian,
			Mamin, and Chui}}]{Rugar2004}
	\bibinfo{author}{\bibfnamefont{D.}~\bibnamefont{Rugar}},
	\bibinfo{author}{\bibfnamefont{R.}~\bibnamefont{Budakian}},
	\bibinfo{author}{\bibfnamefont{H.~J.} \bibnamefont{Mamin}}, \bibnamefont{and}
	\bibinfo{author}{\bibfnamefont{B.~W.} \bibnamefont{Chui}},
	\bibinfo{journal}{Nature} \textbf{\bibinfo{volume}{430}},
	\bibinfo{pages}{329} (\bibinfo{year}{2004}).
	
	\bibitem[{\citenamefont{Tao et~al.}(2014)\citenamefont{Tao, Boss, Moores, and
			Degen}}]{Tao2014}
	\bibinfo{author}{\bibfnamefont{Y.}~\bibnamefont{Tao}},
	\bibinfo{author}{\bibfnamefont{J.~M.} \bibnamefont{Boss}},
	\bibinfo{author}{\bibfnamefont{B.~A.} \bibnamefont{Moores}},
	\bibnamefont{and} \bibinfo{author}{\bibfnamefont{C.~L.} \bibnamefont{Degen}},
	\bibinfo{journal}{Nature Commun.} \textbf{\bibinfo{volume}{5}},
	\bibinfo{pages}{3638} (\bibinfo{year}{2014}).
	
	\bibitem[{\citenamefont{Vinante et~al.}(2016)\citenamefont{Vinante, Bahrami,
			Bassi, Usenko, Wijts, and Oosterkamp}}]{Vinante2016}
	\bibinfo{author}{\bibfnamefont{A.}~\bibnamefont{Vinante}},
	\bibinfo{author}{\bibfnamefont{M.}~\bibnamefont{Bahrami}},
	\bibinfo{author}{\bibfnamefont{A.}~\bibnamefont{Bassi}},
	\bibinfo{author}{\bibfnamefont{O.}~\bibnamefont{Usenko}},
	\bibinfo{author}{\bibfnamefont{G.}~\bibnamefont{Wijts}}, \bibnamefont{and}
	\bibinfo{author}{\bibfnamefont{T.}~\bibnamefont{Oosterkamp}},
	\bibinfo{journal}{Phys.\ Rev.\ Lett.} \textbf{\bibinfo{volume}{116}},
	\bibinfo{pages}{090402} (\bibinfo{year}{2016}).
	
	\bibitem[{\citenamefont{Reinhardt et~al.}(2016)\citenamefont{Reinhardt,
			M\"uller, Bourassa, and Sankey}}]{Reinhardt2016}
	\bibinfo{author}{\bibfnamefont{C.}~\bibnamefont{Reinhardt}},
	\bibinfo{author}{\bibfnamefont{T.}~\bibnamefont{M\"uller}},
	\bibinfo{author}{\bibfnamefont{A.}~\bibnamefont{Bourassa}}, \bibnamefont{and}
	\bibinfo{author}{\bibfnamefont{J.~C.} \bibnamefont{Sankey}},
	\bibinfo{journal}{Phys.\ Rev.\ X} \textbf{\bibinfo{volume}{6}},
	\bibinfo{pages}{021001} (\bibinfo{year}{2016}).
	
	\bibitem[{\citenamefont{Kimble}(2008)}]{Kimble2008}
	\bibinfo{author}{\bibfnamefont{H.~J.} \bibnamefont{Kimble}},
	\bibinfo{journal}{Nature} \textbf{\bibinfo{volume}{453}},
	\bibinfo{pages}{1023} (\bibinfo{year}{2008}).
	
\end{thebibliography}

\clearpage

\section{Supplementary Information}

\subsection{Cooling to Low Phonon Occupancies}

Cooling to the motional ground state of a mechanical oscillator with frequency $\omega_\mathrm{m}$ and linewidth $\Gamma_\mathrm{m}$ (FWHM) using an optical cavity with frequency $\omega_\mathrm{c}$ and amplitude decay rate $\kappa$ in practice means reducing the average thermal occupation to $\bar{n}_\mathrm{f}<1$. In order to resolve the mechanical noise spectrum this also means that the effective mechanical quality factor $Q_{\mathrm{eff}}>1$. $Q_{\mathrm{eff}}$ is the optomechanically broadened quality factor and is different from the intrinsic quality factor $Q_\mathrm{m}=\omega_\mathrm{m}/\Gamma_\mathrm{m}$. To experimentally achieve such cooling performances, we significantly increase the optomechanical coupling from bare Si$_3$N$_4$ resonators by adding photonic crystals, which allow us to reach reflectivities between 95$\%$ and 99$\%$ (see main text). In the following analysis, we calculate the phonon occupancies and the corresponding effective quality factors that are achievable by optomechanically cooling the mechanical resonators presented in this Letter.

\subsubsection{ Effective Mechanical Frequency and Damping Rates}

The optomechanical coupling changes the mechanical properties of the resonator, giving rise to an effective frequency $\omega_{\mathrm{eff}}^2(\omega)=\omega_\mathrm{m}^2+2\omega\omega_{\mathrm{OM}}(\omega)$ and damping rate $\Gamma_{\mathrm{eff}}(\omega)=\Gamma_\mathrm{m}+\Gamma_{\mathrm{OM}}(\omega)$. $\omega_{\mathrm{OM}}(\omega)$ and $\Gamma_{\mathrm{OM}}(\omega)$ are the shifted frequency and damping rate due to the optomechanical interaction, derived from the modified mechanical susceptibility in frequency space $\omega$, which are given by \cite{Aspelmeyer2014}
$$\omega_{\mathrm{OM}}(\omega)=g^2\frac{\omega_\mathrm{m}}{\omega}\left( \frac{\Delta+\omega}{(\Delta+\omega)^2+\kappa^2} + \frac{\Delta-\omega}{(\Delta-\omega)^2+\kappa^2} \right),$$ $$\Gamma_{\mathrm{OM}}(\omega)=g^2\frac{\omega_\mathrm{m}}{\omega}\left( \frac{\kappa}{(\Delta+\omega)^2+\kappa^2} - \frac{\kappa}{(\Delta-\omega)^2+\kappa^2} \right),$$
where $\omega_\mathrm{l}$ is the laser frequency, $\Delta=\omega_\mathrm{l}-\omega_\mathrm{c}$ the detuning with respect to the cavity, $g=g_0\sqrt{\bar{n}_\mathrm{c}}$ the linearized optomechanical coupling strength, $\bar{n}_\mathrm{c}=\frac{2P}{\hbar \omega_\mathrm{l}}\frac{\kappa_1}{\kappa^2+\Delta^2}$ the average cavity photon number, $P$ the laser power, $\kappa_1$ the amplitude decay rate of the input mirror, $\hbar$ the reduced Planck constant, $g_0=-\frac{\partial \omega_\mathrm{c}}{\partial x}x_{\mathrm{zpf}}$ the single-photon optomechanical coupling, $x_{\mathrm{zpf}}=\sqrt{\frac{\hbar}{2m_{\mathrm{eff}}\omega_\mathrm{m}}}$ the zero-point fluctuation of the mechanical mode, and $m_{\mathrm{eff}}$ the effective mass of the mechanical mode. Note that these effects are most pronounced for $\omega=\omega_\mathrm{m}$.

\subsubsection{Final Average Phonon Occupancy}

Using the previous equations one can calculate the effective mechanical quality factor $Q_{\mathrm{eff}}=\frac{\omega_{\mathrm{eff}}}{\Gamma_{\mathrm{eff}}}$ and the oscillator's final average phonon occupancy, which in general is given by
$$\bar{n}_\mathrm{f} = \frac{\Gamma_{\mathrm{OM}}\bar{n}_{\mathrm{min}}+\Gamma_\mathrm{m} \bar{n}_{\mathrm{th}}}{\Gamma_{\mathrm{OM}}+\Gamma_\mathrm{m}}$$
where $\bar{n}_{\mathrm{th}}=(e^{\frac{\hbar \omega_\mathrm{m} }{k_\mathrm{B}T}}-1)^{-1}$ is the average phonon occupancy of a thermal bath at temperature $T$, $k_\mathrm{B}$ is Boltzmann's constants, and $\bar{n}_{\mathrm{min}}=\left( \frac{\kappa^2+(\Delta-\omega_\mathrm{m})^2}{\kappa^2+(\Delta+\omega_\mathrm{m})^2} -1\right)^{-1}$ is the minimum phonon occupancy.

Note that $g_0=-\frac{\partial \omega_\mathrm{c}}{\partial x}x_{\mathrm{zpf}}$ depends on the configuration of the cavity through $\omega_\mathrm{c}(x)$. We estimate $\bar{n}_\mathrm{f}$ and $Q_{\mathrm{eff}}$ starting from room temperature for two different configurations of particular interest, namely the 2 Mirror (2MC) and the 3 Mirror (3MC) Configurations, described below. For both configurations we use the parameters indicated in table~\ref{table:2}.

\begin{table}
	\begin{tabular}{ |c | c | c | c | c | c|}
		\hline
		$~~~\lambda_\mathrm{c}$, $\lambda_l~~~$	& $~~~L~~~$	& $~~\omega_\mathrm{m}/2\pi~~$ & $~~\Gamma_\mathrm{m}/2\pi~~$	& $~~m_{\mathrm{eff}}~~$& $~~\Delta$ \\ \hline
		$1550\,\mathrm{nm}$		& $100\,\mathrm{mm}$	& $140\,\mathrm{kHz}$	& $1.4\,\mathrm{mHz}$		& $1\,\mathrm{ng}$		& $-\omega_\mathrm{m}$ \\ \hline
	\end{tabular}
	\caption{Nominal parameters for cooling estimates.}
	\label{table:2}
\end{table}

\subsubsection{Two Mirror Configuration (2MC)}

In a 2MC, the cavity is composed of two mirrors, in which one of them is the mechanical resonator. In this case, $$\omega_\mathrm{c}^{\mathrm{2MC}}(x)=n\frac{c}{2x},$$ where $n$ is the index of the cavity resonance of interest. If we let one of the mirrors be a distance $x$ away from a predefined cavity length $L$, $\omega_\mathrm{c}(x)=n\frac{c}{2(L+x)}$. For small $x$ we get $-\frac{\partial \omega_\mathrm{c}}{\partial x}= -n\frac{c}{2L^2}=\frac{\omega_\mathrm{c}(L)}{L}$. Simplifying the notation $\omega_\mathrm{c}(L)$ by $\omega_\mathrm{c}$ we get $$g_{0}^{\mathrm{2MC}} = \frac{\omega_\mathrm{c}}{L}\sqrt{\frac{\hbar}{2m_{\mathrm{eff}}\omega_\mathrm{m}}}.$$

We now calculate the achievable average phonon occupancies and mechanical quality factors from cooling membranes such as the ones presented in this Letter, in a configuration where the membranes are used as input mirrors $R_{\mathrm{PhC}}$ in a 2MC in combination with a second mirror with reflectivity $R_2$. Taking $R_2=99.995\%$ and $P=10\,\mathrm{\mu W}$, we show the estimated $\bar{n}_\mathrm{f}$ and $Q_{\mathrm{eff}}$ for several values of $R_{\mathrm{PhC}}$ in table~\ref{table:3}. Here $C_0=4g_0^2/\kappa\Gamma_{\mathrm{m}}$ is the single-photon cooperativity and $2\kappa/\omega_\mathrm{m}$ is the sideband resolution parameter. While in such a configuration it is not possible to enter the ground state regime with our currently best reflectivities of $99.3\%$, future minor improvements in fabrication and design will allow us the reach the required reflectivities of around $99.9\%$. In addition, as is shown in the next section, a membrane in the middle approach (3MC) allows us to realistically cool to the quantum mechanical ground state with our current devices.

\begin{table}
	\begin{tabular}{| c | c  c  c  c |}
		\hline
		$~~R_{\mathrm{PhC}}~~$ 	& $~2\kappa/\omega_\mathrm{m}~$ & $~C_0~$ & $~\bar{n}_\mathrm{f}~$	& $~Q_{\mathrm{eff}}~$ \\ \hline
		95\%	& 85			& 0.1	&	$1.8\times10^5$ & $4.1\times10^5$\\ \hline
		99.3\%	& 12		& 0.8		& 90	& 188\\ \hline
		99.9\%	& 1.8		& 5.1		&	0.7 & 1.4 \\ \hline
	\end{tabular}
	\caption{Estimated cooling performance for a 2MC, assuming $R_2=99.995\%$ and $P=10\,\mathrm{\mu W}$.}
	\label{table:3}
\end{table}

\subsubsection{Three Mirror Configuration (3MC)}

In a 3MC, the membrane is placed at an arbitrary position $x$ between two high highly reflective mirrors which are separated by a fixed distance $L$. In this case~\cite{jayich2008dispersive}, $$\omega_\mathrm{c}^{\mathrm{3MC}}(x)=\omega_\mathrm{c}\left[2\phi_{\mathrm{r}}+2\cos^{-1}\left(\left|r_\mathrm{m}\right|\cos\left(\frac{4\pi}{\lambda_\mathrm{c}}x\right)\right)\right]$$
where $r_\mathrm{m}$ is the amplitude reflectance coefficient of the membrane, $\phi_{\mathrm{r}}=\arg(r_\mathrm{m})$, and $\lambda_\mathrm{c}$ is the cavity wavelength. 
Note that $-\frac{\partial \omega_\mathrm{c}(x)}{\partial x}$ is optimal for $\frac{4\pi}{\lambda}x=\frac{\pi}{2}+n\pi$. Assuming the membrane is positioned at such an optima and that it is allowed to move a small distance $\epsilon$ away from the optimum, i.e.\ $\frac{4\pi}{\lambda}x=\frac{\pi}{2}+\epsilon$, it is possible to approximate $\omega_\mathrm{c}(\epsilon)\approx \omega_\mathrm{c}\left[2\phi_{\mathrm{r}}-2\left|r_\mathrm{m}\right|\epsilon\right]$. As such, $-\frac{\partial \omega_\mathrm{c}(x)}{\partial x}=-\frac{4\pi}{\lambda}\frac{\partial \omega_\mathrm{c}(\epsilon)}{\partial \epsilon}=2\left|r_\mathrm{m}\right|\frac{\omega_\mathrm{c}}{L}$ and we get, $$g_{0}^{\mathrm{3MC}} = 2\left|r_\mathrm{m}\right|\frac{\omega_\mathrm{c}}{L}\sqrt{\frac{\hbar}{2m_{\mathrm{eff}}\omega_\mathrm{m}}}.$$

The advantage of this configuration is that the optomechanical mirror with reflectivity $\left|r_\mathrm{m}\right|^2=R_{\mathrm{PhC}}$ can be positioned inside a cavity where the finesse $F$ is only determined by the fixed mirrors and not by the membrane reflectivity itself. In such an approach the influence of the membrane's optical properties on the achievable cooling performance is significantly smaller compared to the case for 2MC. Even when using our design with the lowest reflectivity $R_{\mathrm{PhC}}=95\%$, $P=10\,\mathrm{\mu W}$, we see that occupancies below 1 phonon are easily accessible with a modest finesse cavity. In table~\ref{table:4} the estimated $\bar{n}_\mathrm{f}$ and $Q_{\mathrm{eff}}$ for varying cavity finesse are shown. In particular, for $F=4100$ we estimate $\bar{n}_\mathrm{f}=0.67$ and $Q_{\mathrm{eff}}=1.0$. These calculations show that our novel designs place the quantum ground state of an optomechanical device at room temperature well within experimental reach.

\begin{table}
	\begin{tabular}{| c | c  c  c  c |}
		\hline
		$~~F~~$ 	& $~2\kappa/\omega_\mathrm{m}~$ & $~C_0~$ & $~\bar{n}_\mathrm{f}~$	& $~Q_{\mathrm{eff}}~$ \\ \hline
		3400	& 3.2			& 11	&	0.95 & 1.3\\ \hline
		4100	& 2.6		& 13		& 0.67	& 1.0\\ \hline
		18000	& 0.6		& 58		&	0.10 & 0.3 \\ \hline
	\end{tabular}
	\caption{Estimated cooling performance for a 3MC assuming a PhC reflectivity of $95\%$, the lowest reflectivity achieved by our devices, and $P=10\,\mathrm{\mu W}$.}
	\label{table:4}
\end{table}

\subsection{Fabrication}
Our mechanical resonators are fabricated by depositing a high-stress (1.3~GPa) silicon nitride film in a LPCVD furnace on $\langle100\rangle$ silicon wafers. Tethered membrane patterns written into electron beam lithography resist (AR-P 6200) are then transferred into the silicon nitride film using an ICP-RIE pseudo-Bosch etch (C$_4$F$_8$/SF$_6$) and cleaned with an O$_2$ plasma followed by a hot piranha solution to remove organics. Another layer of electron-beam resist is spun over the (unreleased) trampoline and all parts of the tethered membrane except the central region containing the photonic crystal mirror are then exposed, which allows us to thin the membrane tethers using a plasma etch while keeping a thick central membrane,  resulting in devices with both high reflectivity and low mechanical dissipiation. The devices are then undercut by performing an aligned beamwrite to open a backside window in the silicon nitride. The resonators are subsequently released from the silicon substrate using $30\%$ potassium hydroxide (KOH) at 75~$^{\circ}$C. After the release, we treat the ultrathin resonators with extreme care by avoiding their exposure to large temperature gradients, viscous forces and surface tension using PTFE turbuluence shielding holders. As the resonators decrease in thickness, large tensile stress concentrates in ever smaller clamping cross sections.

\subsection{Resonator Design}
Our devices are designed and optimized by performing a systematic analysis of several parameters of the tethered membranes (see Figure~1, main text) in order to enhance the $f \times Q_\mathrm{m}$ product. Parameters we studied are the size of the central membrane ($a_{\mathrm{mem}}$), width of the tethers ($w$), total size of the window ($a_\mathrm{w}$), fillet radius connecting the tethers to the central membrane ($r_{\mathrm{in}}$), fillet radius connecting the tethers to the substrate ($r_{\mathrm{out}}$), diameter of the photonic crystal mirror (PhC) ($d_{\mathrm{PhC}}$), Si$_3$N$_4$ film thickness ($t_\mathrm{f}$) and silicon substrate thickness ($t_\mathrm{s}$) to measure their effects on the $f \times Q_\mathrm{m}$ product of the center-of-mass mode.

\begin{table}[h!]
	\centering
	\begin{tabular}{| c | c | c | c | } 
		\hline
		& $f$ & $Q_\mathrm{m}$ & $f \times Q_\mathrm{m}$ \\ [0.5ex] 
		\hline\hline
		tether width $w,\uparrow$ & $~~~~~\uparrow~~~~~$ & $~~~~~\downarrow~~~~~$ & $~~~~~=~~~~~$ \\ 
		window size $a_{\mathrm{w}}, \uparrow$ & $\downarrow$ & $\uparrow$ & $=$ \\
		~~membrane size $a_{\mathrm{mem}}, \uparrow$~~ & $\downarrow$ & $\uparrow$ & $\downarrow$ \\
		inner fillet $r_{\mathrm{in}}, \uparrow$ & $\downarrow$ & $\uparrow$ & $=$\\
		outer fillet $r_{\mathrm{out}}, \uparrow$ & $\uparrow$ & $\uparrow$ & $\uparrow$ \\ [1ex] 
		PhC diameter $d_{\mathrm{PhC}}, \uparrow$ & $=$ & $=$ & $=$ \\
		\hline
	\end{tabular}
	\caption{Using 50~nm thick Si$_3$N$_4$ films on 500~$\mu$m silicon substrates as a standard platform, we sweep all design parameters to determine their effect on $f$ and $Q_\mathrm{m}$. Here $\uparrow$ means an increase, while $\downarrow$ a decrease of the associated value. Most parameters suffer from a trade-off between $f$ and $Q_\mathrm{m}$, resulting in negligible enhancement ($=$).}
	\label{table:X}
\end{table}

\begin{figure}[h]
	\begin{center}
		\includegraphics[width=1\columnwidth]{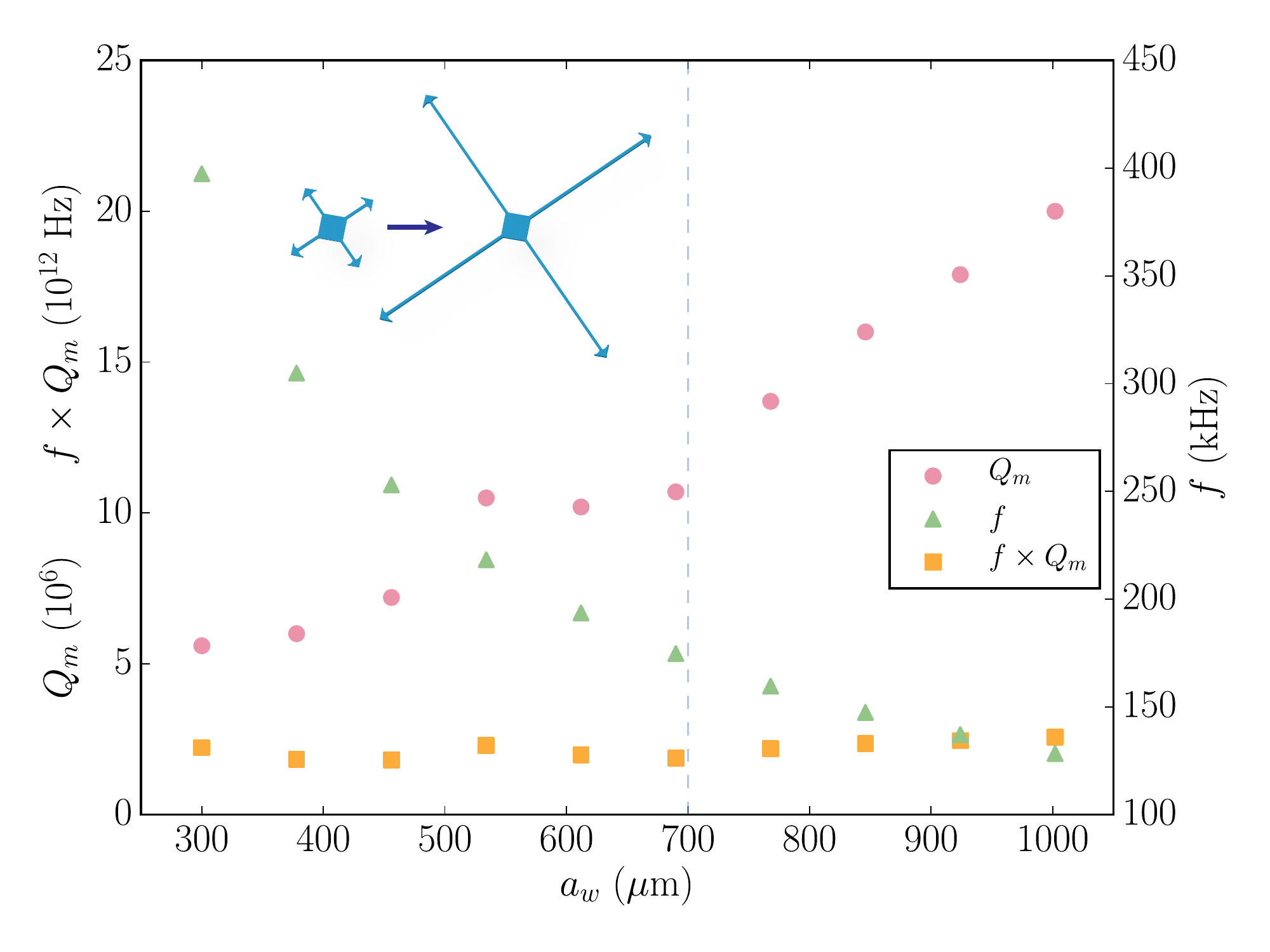}
		\caption{Sweep of the window size $a_\mathrm{w}$.}
		\label{fig:SI1}
	\end{center} 
\end{figure}

In order to measure the effects of the different design parameters on the mechanical quality factor and frequency we start with a set of nominal parameters and sweep each one while keeping all other parameters fixed. Our nominal parameters are: $w=5~\mu$m, $a_\mathrm{w}=700~\mu$m, $a_\mathrm{mem} = 100~\mu$m, $r_\mathrm{in} = 60~\mu$m, $r_\mathrm{out} = 20~\mu$m. The following plots include a blue-dashed line representing the nominal value of the swept parameter. By changing the size of the inside and outside fillets we effectively change the length of the tethers. In order to decouple these parameters we change the size of the window, $a_\mathrm{w}$, as a way to change the length of the tethers. While the relationship of these parameters with frequencies is experimentally robust, one of the main challenges is observing these parameters' effects on the mechanical quality factors which are susceptible to unavoidable fabrication imperfections due to small changes in handling and contaminants from sample to sample. To mitigate this effect, all resonators for these sweeps are fabricated with $t_\mathrm{f} = 50$~nm and $t_\mathrm{s} = 500~\mu$m, where we know that our quality factors are consistently high. The first parameter we investigate is the length of the tethers which we effectively vary by changing the size of the square window ($a_\mathrm{w}$). As shown in Figure~\ref{fig:SI1}, increasing the length of the tethers increases $Q_\mathrm{m}$ and decreases $f$, as expected, with almost no change in the $f \times Q_\mathrm{m}$ product. 

\begin{figure}[h]
	\begin{center}
		\includegraphics[width=1\columnwidth]{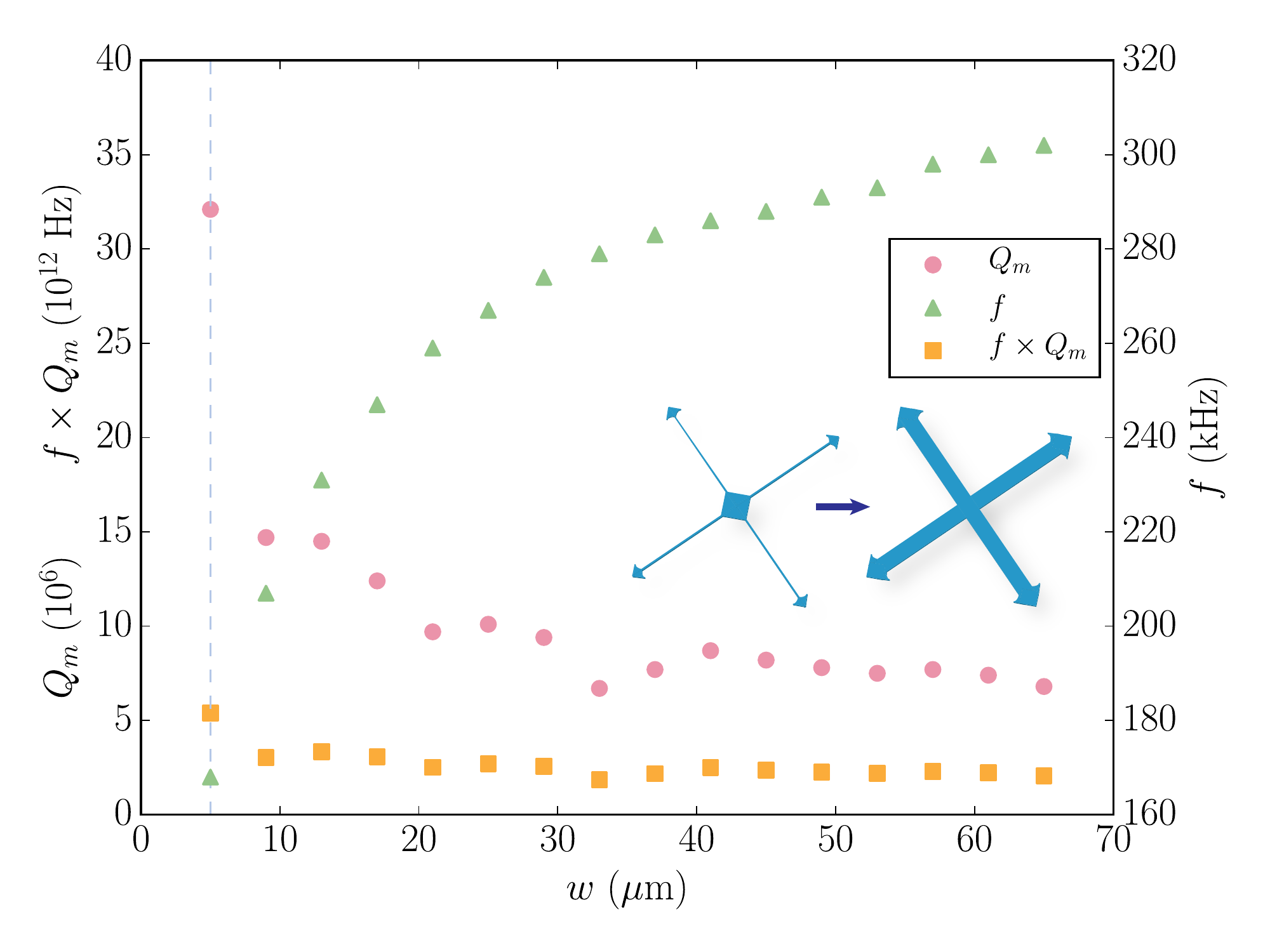}
		\caption{Sweep of tether width $w$.}
		\label{fig:SI2}
	\end{center} 
\end{figure}

We next sweep the width of the tethers from the nominal value of 5~$\mu$m to 64~$\mu$m, which is the maximum width before the tethers begin to overlap (Figure~\ref{fig:SI2}). We measure the mechanical frequency increase and quality factor decrease towards a plateau near $7 \times 10^{6}$, which is also the quality factor for a 50~nm thick square membrane of 700~$\mu$m. Intuitively in the limit of increasing tether width the tethered membranes begin to resemble square membranes where the tether width can no longer be properly defined and only the size of the window, $a_\mathrm{w}$, defines the lateral geometry of the membrane. We see a large enhancement in $Q_\mathrm{m}$ from $15 \times 10^6$ to $32 \times 10^6$ when we decrease the tether width from 9~$\mu$m to 5~$\mu$m as the membranes losses begin to be dominated by clamping losses associated with strings ($Q_{\mathrm{clamp}}\propto a_\mathrm{w}/w$).

\begin{figure}[h]
	\begin{center}
		\includegraphics[width=1\columnwidth]{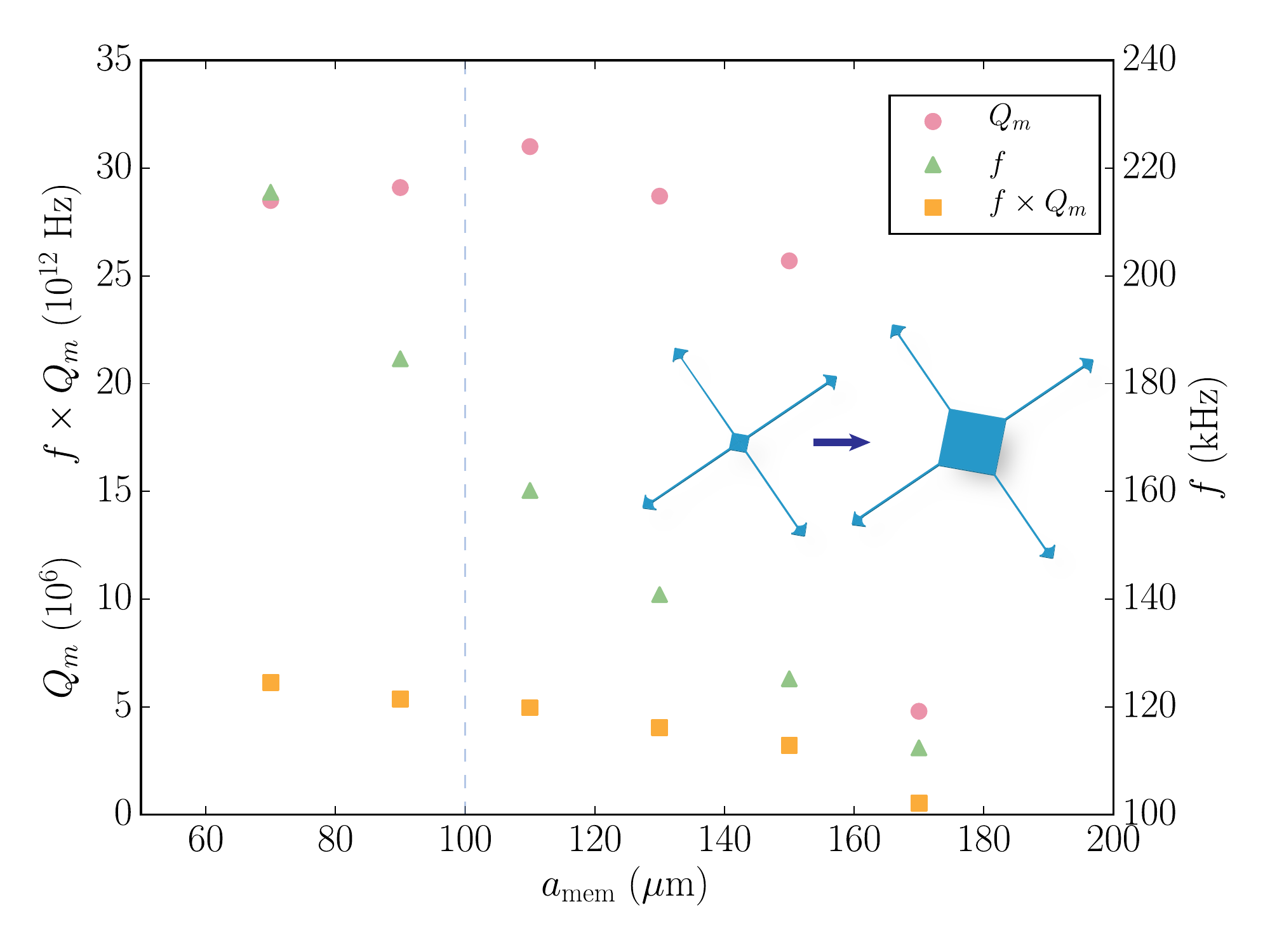}
		\caption{Sweep of the membrane size $a_\mathrm{mem}$.}
		\label{fig:SI3}
	\end{center} 
\end{figure}

From finite element simulations we know that the contraction of relaxing material in the central membrane plays a key role in the enhancement of stress in the tethers. We sweep the size of the central membrane, $a_\mathrm{mem}$, in order to see how it enhances the stress in the tethers and ideally increases $f \times Q_\mathrm{m}$. As we increase $a_\mathrm{mem}$, we simultaneously also increase $a_\mathrm{w}$ to keep the length of the tethers constant. We find that the mechanical quality factors remain at nominal values while the frequency drops dramatically as the resonator mass increases quadratically (Figure~\ref{fig:SI3}). In this case, $f \times Q_\mathrm{m}$ improves with decreasing membrane size. From a practical standpoint, the optimal size of the membrane is set by the minimum cavity beam waist which can still achieve large reflectivities from the photonic crystal on the central membrane. By using a small beam waist the size of the central membrane can be minimized in order to increase $f \times Q_\mathrm{m}$, with the limit of good mode overlap with the photonic crystal to achieve high reflectivities. A beam waist of 20~$\mu$m gives us maximal reflectivity and requires a minimum photonic crystal mirror with $d_{\mathrm{PhC}} \sim 80~\mu$m in order to overlap the tails of the Gaussian beam.

\begin{figure}[h]
	\begin{center}
		\includegraphics[width=1\columnwidth]{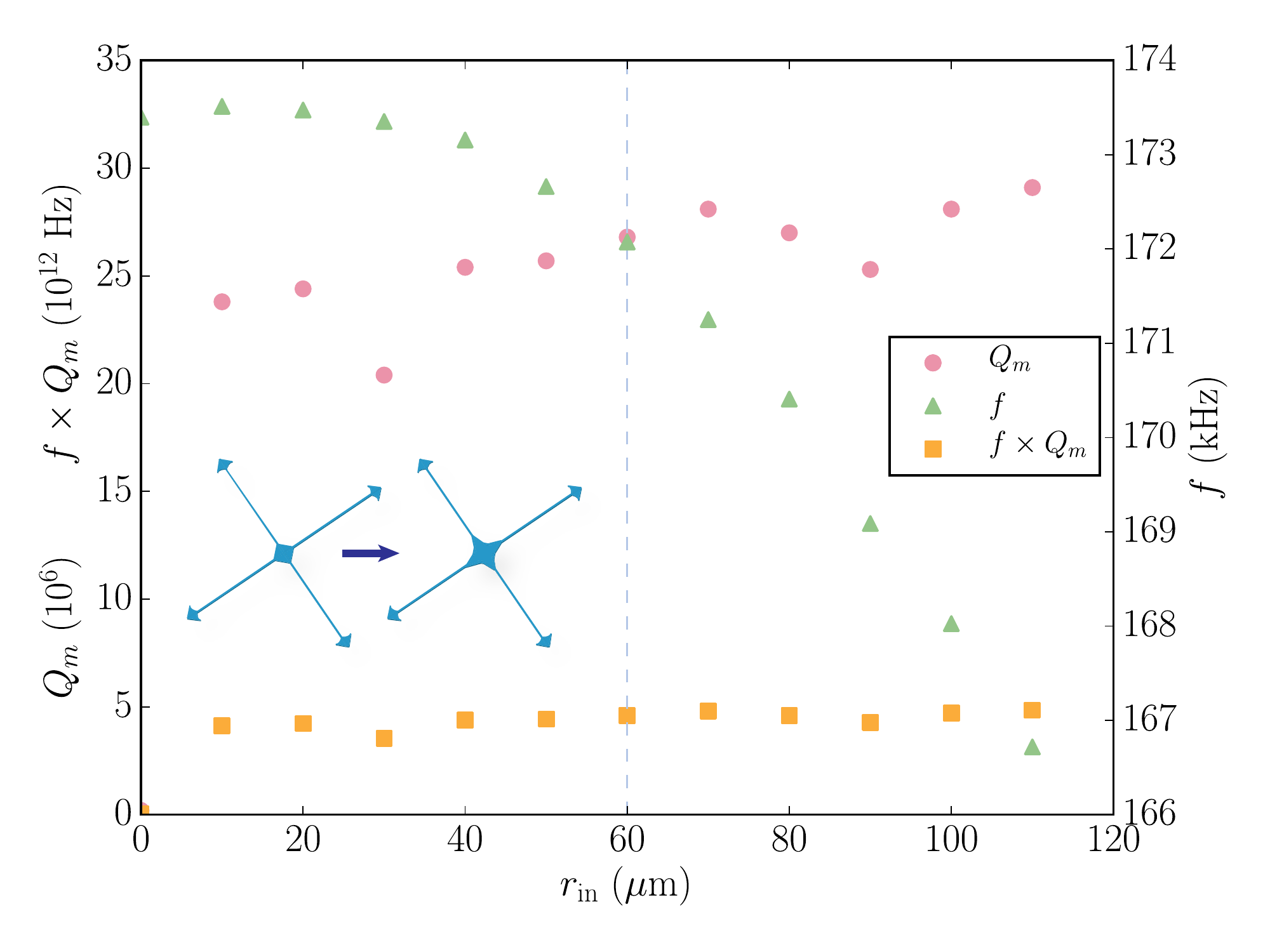}
		\caption{Sweep of the inner fillet $r_\mathrm{in}$.}
		\label{fig:SI4}
	\end{center} 
\end{figure}

In addition, as the inner fillet radius between tethers and central membrane is increased, there is a slight increase in quality factor with a large decrease in frequency as the mass of the central membrane increases (Figure~\ref{fig:SI4}).

\begin{figure}[h]
	\begin{center}
		\includegraphics[width=1\columnwidth]{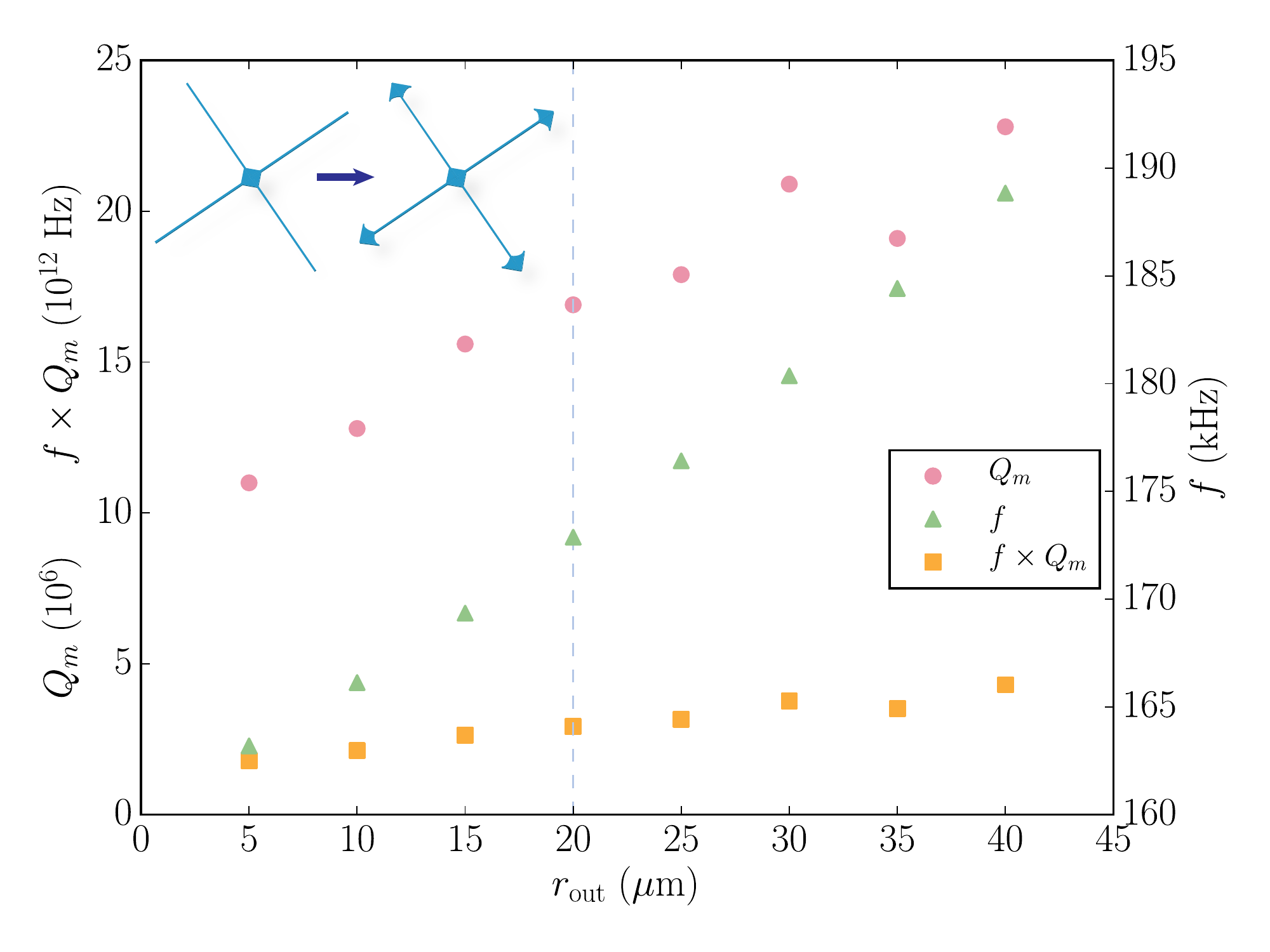}
		\caption{Sweep of the outer fillet $r_\mathrm{out}$.}
		\label{fig:SI5}
	\end{center} 
\end{figure}

We also vary the size of the outer fillet radius between the tether and the edge of the window. It is however difficult to directly compare resonators with different outer-fillet shapes since larger outer-fillets require longer KOH release times. The increased etching times also widens the overhang around the perimeter of the window, which we find to lower the mechanical quality factors for nominal dimensions (Figure~\ref{fig:SI5}).

\begin{figure}[h]
	\begin{center}
		\includegraphics[width=1\columnwidth]{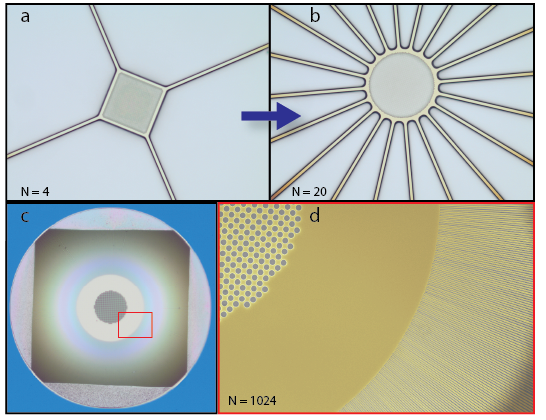}
		\caption{Sweep of the number of tethers $N$.}
		\label{fig:SI6}
	\end{center} 
\end{figure}

Last but not least, we also vary the number of tethers connecting the central membrane to the substrate, $N$. We sweep it from $N=4$ to $N=20$ as shown in Figure~\ref{fig:SI6}(a,b) while keeping all other parameters nominal. The observed shift in frequency in these "pinwheel" flexural resonators is from 170~kHz ($N=4$) to almost 350~kHz ($N=20$). In order to keep the lengths of the tethers equal, a rotationally symmetric design is required. In Figure~\ref{fig:SI6}(c,d), we show a resonator with $N = 1024$ tethers, each 500~nm wide, and a square hole through the substrate for optical access. This large-$N$ design exhibits frequencies around 1~MHz, with $Q_\mathrm{m}$$\sim$$10^6$, which could be an interesting approach to further mitigate classical laser noise. An important fabrication challenge for such devices is to keep the overhang at the perimeter of the resonator to a minimum in order to reduce radiative losses into the substrate. The smallest overhang we were able to attain was $\approx 8~\mu$m wide, which incurred significant losses resulting in $Q_\mathrm{m} \sim 1$--$3 \times 10^6$. The nominal design ($N=4$) keeps the overhang to a minimum due to the nature of rectangular crystal lattice etching of $\langle100\rangle$ silicon.

\subsection{Photonic Crystal Thickness}

\begin{figure}[h]
	\begin{center}
		\includegraphics[width=1\columnwidth]{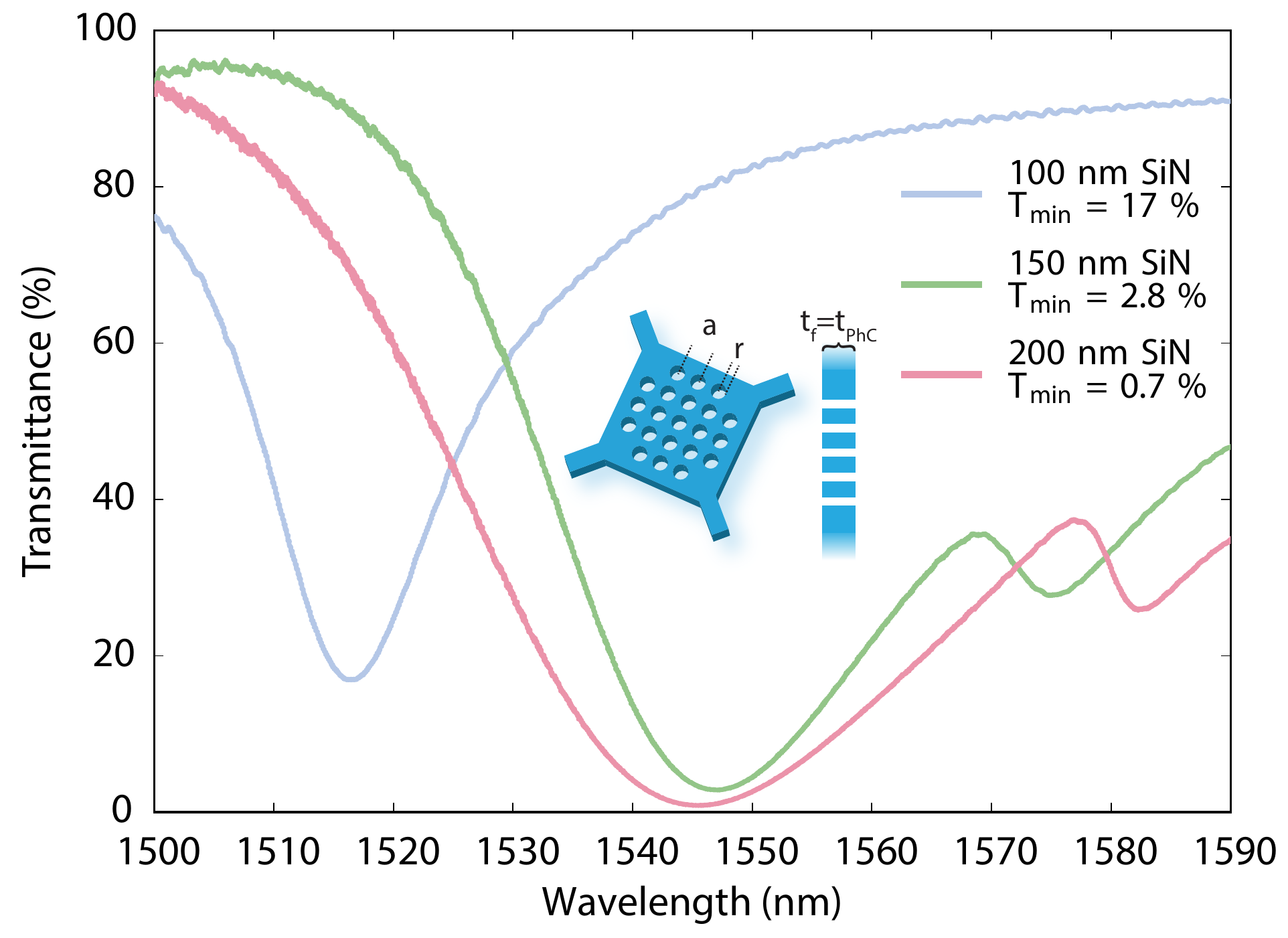}
		\caption{The measured transmittance $T$ of the tethered mechanical membranes as a function of silicon nitride thickness for a standard photonic crystal design of an array of holes (Figure~2b).}
		\label{fig:SI7}
	\end{center} 
\end{figure}

Using FDTD simulations we are able to optimize for the most reflective design at different photonic crystal thicknesses ($t_\mathrm{PhC} = 100,150,200$~nm) for 1550~nm light. In our simulations we sweep the lattice constant $a$ and $r/a$, where $r$ is the radius of the holes. Using these parameters, we are able to experimentally verify the FDTD results which show increased maximum reflectivity with increasing thickness of the photonic crystal $t_\mathrm{PhC}$ (see Figure~\ref{fig:SI7}). While we only plot transmittance $T$ measurements for clarity, we also experimentally confirm that the reflectivity $R=1-T$ to within our measurement error by using the membranes as part of a Fabry-P\'erot cavity in conjunction with another mirror with known reflectivity.

\end{document}